\documentclass[12pt,preprint]{aastex}
\usepackage{epsf}
\usepackage{lscape}
\usepackage{graphicx}
\usepackage{natbib}
\begin{document}
\title{SHELS: Complete Redshift Surveys of Two Widely Separated Fields}
\author {Margaret J. Geller} 
\affil{Smithsonian Astrophysical Observatory,
\\ 60 Garden St., Cambridge, MA 02138}
\email{mgeller@cfa.harvard.edu}
\author {Ho Seong Hwang} 
\affil {School of Physics, Korea Institute for Advanced Study,
\\85 Hoegiro, Dongdaemun-gu, Seoul 130-722, Republic of Korea}
\email{hhwang@kias.re.kr}
\author {Ian P. Dell'Antonio}
\affil{Department of Physics, Brown University,
\\ Box 1843, Providence, RI 02912} 
\email{ian@het.brown.edu}
\author{Harus Jabran Zahid}
\affil{Harvard-Smithsonian Center for Astrophysics,
\\ 60 Garden St., Cambridge, MA 02138}
\email{zahid@cfa.harvard.edu}
 \author {Michael J. Kurtz} 
\affil{Smithsonian Astrophysical Observatory,
\\ 60 Garden St., Cambridge, MA 02138}
\email{mkurtz@cfa.harvard.edu}
\author {Daniel G. Fabricant} 
\affil{Smithsonian Astrophysical Observatory,
\\ 60 Garden St., Cambridge, MA 02138}
\email{dfabricant@cfa.harvard.edu}

\begin{abstract}
The SHELS (Smithsonian Hectospec Lensing Survey) is a complete redshift survey covering two well-separated fields (F1 and F2)  of the Deep Lens Survey. Both fields are more than 94\% complete to a Galactic extinction corrected R$_0$ = 20.2. Here we describe the redshift survey of the F1 field centered at R.A.$_{2000}$ = 00$^h$53$^m$25.3$^s$ and 
Decl.$_{2000}$ = 12$^\circ$33$^\prime$55$^{\prime\prime}$; like F2, the F1 field covers $\sim$4 deg$^2$. The redshift survey of the F1 field includes 9426 new galaxy redshifts measured with Hectospec on the MMT (published here). As a guide to future uses of the combined survey we compare the mass metallicity relation and the distributions of D$_n$4000 as a function of stellar mass and redshift for the two fields. The mass-metallicity relations differ by an insignificant 1.6$\sigma$. For galaxies in the stellar mass range
10$^{10}$ to 10$^{11}$M$_\odot$, the increase in the star-forming fraction with redshift is remarkably similar in the two fields. The seemingly surprising 31-38\% difference in the overall galaxy counts in F1 and F2 is probably consistent with the expected cosmic variance given the subtleties of the relative systematics in the two surveys.  We also review the Deep Lens Survey cluster detections in the two fields: poorer photometric data for F1 precluded secure detection of the single  massive cluster at $z = 0.35$ that we find in SHELS.  Taken together the two fields include 16,055 redshifts for galaxies with R$_0 \leq 20.2$ and 20,754 redshifts for galaxies with R$\leq$ 20.6. These
dense surveys in two well-separated fields provide a basis for future investigations
of galaxy properties and large-scale structure. 

\end{abstract}
\section {Introduction}
Redshift surveys and weak lensing maps are two powerful and independent tools for tracing  the matter distribution in the universe. Approaches to combining these two powerful tools are
developing rapidly as dense redshift surveys access the intermediate redshift universe and as
weak lensing maps become increasingly extensive (Geller et al. 2005; Kurtz et al. 2012; Shan et al. 2012; van Waerbeke et al. 2013; Chang et al. 2015).

The SHELS (Smithsonian Hectospec Lensing Survey) project began as  a platform for exploring the combination of dense, complete foreground redshift surveys with  lensing maps (Geller et al. 2005; Geller et al. 2010; Utsumi et al. 2014). SHELS consists of dense redshift surveys of two of the five fields of the Deep Lens Survey (Wittman et al. 2006; DLS), F1 centered at  R.A.$_{2000}$ = 00$^h$53$^m$25.3$^s$, Decl.$_{2000}$ = 12$^\circ$33$^\prime$55$^{\prime\prime}$ and F2 centered at R.A.$_{2000}$ = 09$^h$19$^m$32.4$^s$ and Decl.$_{2000}$ = +30$^{\circ}$00$^{\prime}$00$^{\prime\prime}$. Each of the DLS fields covers 4 square degrees. The SHELS redshift surveys of the two fields are more than 94\% complete to an extinction corrected R$_0$ = 20.2. These two fields currently represent the most densely sampled surveys to this magnitude limit. The dense, complete sampling makes the surveys useful for a wide range of astrophysical applications.

The Hectospec, a 300-fiber spectrograph with a 1$^\circ$ field of view mounted on the MMT, enables surveys like SHELS. Kochanek et al. (2012) also carried out a large redshift survey, AGES, with Hectospec. Their survey focuses on AGN evolution. The AGES survey covers a 7.7$^\circ$ contiguous region with a complex sampling strategy in several photometric bands. To a
extinction corrected limit of R$_0$ = 20.2 (we use the subscript $_0$ throughout to denote an extinction corrected magnitude), the number density of AGES redshifts is $\sim$1350 galaxies deg$^{-2}$ in contrast with the SHELS mean averaged over the two fields of 1961 galaxies deg$^{-2}$. SHELS and AGES are thus complementary
in both their geometry  and selection.

Here we describe the SHELS survey of the F1 DLS field. Geller et al. (2014) describe the survey of the F2 field. The F1 field contains a total of 9426 new galaxy redshifts and is 94\% complete
to an extinction corrected R$_0$ = 20.2. The F2 survey is somewhat deeper; it is 95\% complete to an observed R = 20.6 and 97\% complete to R$_0$ = 20.2. Taken at face value, the two fields seem remarkably different; the raw counts to R$_0$ = 20.2 differ by 
$\sim$31\%. The F2 field contains a prominent complex of rich clusters associated with Abell 781 (Abell 1958) and easily detected with weak lensing; F1 contains
no clusters detected with weak lensing in the entire volume covered by the redshift survey
(Ascaso et a. 2014). We compare these observed differences with the expected cosmic variance for these fields. We also provide a guide to the properties of the survey and to some of its potential uses.

We have already used both the F2 and F1 SHELS data for a variety of applications. For essentially all of these applications, the  straightforward, complete magnitude limited survey is a critical underpinning. In concert with the original intent of the surveys, we have used the F2 survey as a basis for testing weak lensing maps against a foreground redshift Survey (Geller et al. 2005; Geller et al. 2010; Utsumi et al. 2014; see also
Viola et al. 2015). We have also compared x-ray, spectroscopic and lensing selection
demonstrating that the most massive systems are detected robustly with all three techniques (Starikova et al. 2014). 

The extensive SHELS spectroscopy enables exploration of a variety of benchmarks for 
tracking galaxy evolution including the evolution of the H$\alpha$ luminosity function (Westra et al. 2010), the faint end of the composite galaxy luminosity function (Geller et al. 2012), the nature of star-forming galaxies detected with WISE (Hwang et al. 2012),  the impact of close pairs on star formation (Freedman Woods et al. 2010),  the determination of central velocity dispersions for individual galaxies (Fabricant et al. 2013), and the evolution of the mass-metallicity relation (Zahid et al. 2013; 2014). Here we provide guides to the quality and potential uses of  the SHELS surveys by revisiting the stellar mass metallicity relation; we compare the determinations for the F1 and F2 fields separately and in combination.
Similarly, we compare the D$_n$4000 distributions as a function of stellar mass and redshift for the
two fields.

We describe the F1 data in Section 2 with attention to the differences between the photometric bases for the F1 and F2 surveys. We compare the completeness of the shallower F1 survey with F2 and lay the foundation for comparing the surveys to the same limiting extinction corrected apparent magnitude. In Section 3 we compare various aspects of the F1 and F2 survey regions including  the mass metallicity relation (Section 3.1), the D$_n$4000 distributions as a function of redshift and stellar mass (Section 3.2), galaxy counts (Section 3.3), and cluster observations (Section 3.4). In Sections 3.3 and 3.4 we consider whether the differences in galaxy and cluster counts are consistent with the expected cosmic variance for fields of this size. We conclude in Section 4.

We adopt   H$_0$ = 70 km s$^{-1}$ Mpc$^{-1}$, $\Omega_{\Lambda} = 0.7$ and $\Omega_m$ = 0.3 throughout. All quoted errors in measured quantities are 1$\sigma$.

\section {The Data}

The SHELS redshift survey covers  two 4 deg$^2$ fields originally selected as part of the
Deep Lens Survey (DLS; Wittman et al 2006). Like most fields chosen for deep photometric surveys, the selection of the Deep Lens Survey fields avoids bright stars, nearby bright galaxies, and nearby rich clusters with redshift $z \lesssim 0.1$. This selection biases the fields toward low density at $z \lesssim 0.1$. 

Geller et al. (2014) describe the redshift survey of the F2 field of the DLS. In this field, the redshift survey is 95\% complete to R = 20.6  (observed total magnitude  uncorrected for Galactic extinction). The complete portion of the survey includes 12,705 redshifts. It is interesting to note that the F2 field contains an impressive complex of rich clusters at z $\sim$ 0.3 and another at z $\sim$ 0.5 (Geller et al. 2010; Ascaso et al. 2014; Utsumi et al. 2014; Starikova et al. 2014).

The original intent of SHELS was to complete the redshift survey of the F1 field to R = 20.6.
Although we measured many redshifts for galaxies to this limit (and we report them here), observing conditions only allowed a survey 93\% complete to an extinction corrected R$_0$ = 20.2. Furthermore the DLS 
reported no significant weak lensing peaks in F1 (see Ascaso et al. 2014) thus diminishing the incentive 
for further spectroscopy (but see the discussion of clusters in Section 3.4). 

For the F1 field $A_R$ = 0.16 in contrast with $A_R$ = 0.05 for F2. In the discussion below we compare the properties of the two fields at the  limiting R$_0$ = 20.2. To this limit, the F2 field contains 9489 galaxies (9216 of these have redshifts) ; remarkably, the F1 field contains 
only 7261 galaxies (6839 of these have redshifts). Together the two fields currently constitute the largest complete redshift surveys to this limit. For comparison VVDS-Wide is deeper (I$_{AB} <$ 22.5) than SHELS
but sparsely sampled (at the 20-25\% level); the
target density is ~4800 deg$^{-2}$ and it covers a similar area (Garilli
et al. 2008). The GAMA survey is somewhat shallower (r $< $19.8) and has a
target density about a factor of 2 smaller than SHELS, but it covers
a much wider area, and is  more complete (Liske et
al. 2015).

\subsection {Photometry}
\label{photometry}
Geller et al. (2014) describe the construction of the F2 catalog directly from the DLS R-band photometry (Wittman et al. 2006).
For this field we masked out regions around bright stars (equation (1) in Geller et al. 2014), thus reducing the effective survey area to 3.98 square degrees.  

\begin{figure}
\centerline{\includegraphics[width=7.0in]{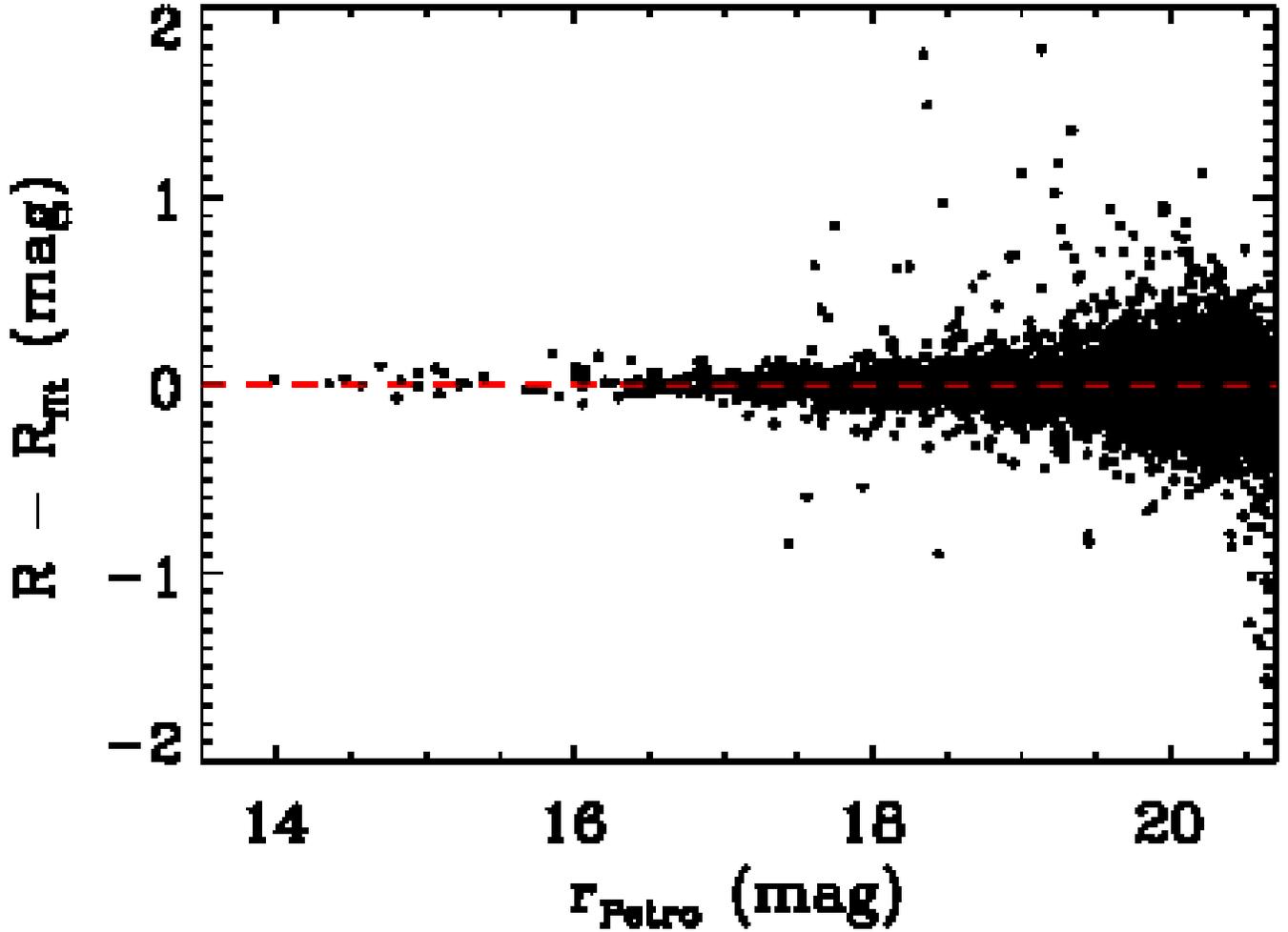}}
\vskip 5ex
\caption{Residuals, R-R$_{fit}$, between the DLS R-band F2 magnitude (R) and R$_{fit}$ (equation 1) as a function of the SDSS r-band Petrosian magnitude, r$_{petro}$. The typical magnitude errors
are 0.01 (DLS R) and 0.11 (SDSS r$_{petro}$).  
\label{magcomp}}
\end{figure}

In the F1 field we began by selecting targets from the SDSS DR9 (Ahn et al. 2012) in a region somewhat larger than the DLS F1 footprint.  The main galaxy candidates are extended sources based on the SDSS point/extended source flag.  We do not used the point sources in the analysis; we  use only the extended sources.  We then use the available DLS data to construct an R-limited galaxy catalog. For each SDSS galaxy candidate with r$_{petro} < 23$, we searched for a DLS counterpart within a 2$^{\prime\prime}$ tolerance. If there is a unique DLS counterpart, we adopt the DLS R-band magnitude. If there is no DLS counterpart (generally as a result of a nearby saturated star, a bleed trail, or proximity to the DLS survey edges),
we transform the magnitudes to the same system we use in F2 (Geller at al. 2014). We base the transformation on a fit to all of the galaxies we identify in both the DLS F2 field and the SDSS. We fit the DLS F2 R-band magnitude (R) as a function of SDSS $r$ and $(g-r)$ color to derive the transformation
\begin{equation}
R_{fit} = r-0.070(g-r)-0.227. 
\end{equation}
Figure \ref{magcomp} shows the residuals between the DLS R-band magnitude and $R_{fit}$
as a function of the SDSS $r_{petro}$.
The total number of objects with R$_0 \leq 20.2$ that require this conversion is
710 or 9.8\% of the sample to this limit. Most of these objects are near the boundaris of the F1 redshift survey region and thus just outside the DLS F1 footprint.

There is also a small offset between the DLS F2 magnitudes provided in their on-line database and the magnitudes
reported in Geller et al (2014). We note that the F2 magnitudes reported in Geller et al. (2014) are total magnitudes calibrated to the Vega system and extrapolated from the isophotal magnitude within the 28.7 mag arcsec$^{-2}$ isophote. The zero point offset between these magnitudes and those provided on the DLS website is 0.038 magnitudes. We add this small offset to magnitudes for galaxies with  F1 photometry  in order to put the two fields on the same system for
a direct comparison.

For F1 we do not mask regions around bright stars;
the effective area covered by the survey is 4.2 square degrees. There may be some undersampling of the galaxy distribution around bright stars, but based on the area excised in F2, a much deeper survey than the SDSS, we expect the effect to be substantially less than 5\%. 

In the F2 field approximately 3\% of the galaxy candidates in our original catalog turned out to be stars. In the F1 field, a comparable 4\% of the galaxy candidates selected from the SDSS turned out to be stars. The smaller fraction of stars in F2 is qualitatively consistent with better seeing for the DLS imaging.

The DLS photometry is much deeper than the SDSS. Furthermore the seeing  in F2 was 
0.90$^{\prime\prime}$ with a $\sim5$\% variation among subfields; for F1 the average seeing was 0.98$^{\prime\prime}$ varying from 0.94$^{\prime\prime}$ to 1.01$^{\prime\prime}$.
For the SDSS, the median seeing (for DR7) was 1.43$^{\prime\prime}$.
Thus the DLS  photometry may reveal low surface brightness objects and/or objects that are unresolved in the SDSS. 
Remarkably,  the number of F2 galaxies with R$_0 \leq 20.2$ missed in the SDSS photometry is only 37, corresponding to 0.4\% of total number of F2 galaxies with R$_0 \leq$20.2. These objects are indeed either low surface brightness galaxies or they have very close neighbor galaxies unresolved by the SDSS. We may be missing a comparable fraction of galaxies in F1  because we constructed the catalog directly from the SDSS; we conclude that the impact of differences in catalog construction  on the numbers of objects is negligible at this limiting magnitude.   We note that failure to include low surface brightness objects 
is a potentially important limitation for investigating some issues including the faint end of the luminosity function (Blanton et al. 2005; Geller et al. 2012).
\subsection {Spectroscopy}
\label{spectroscopy}
We used the 300-fiber Hectospec instrument (Fabricant et al. 1998, 2005) on the MMT to acquire spectroscopy for galaxy candidates typically brighter than R = 20.6. We observed the F1 field in queue mode during dark runs in four periods:
October 24 - 28, 2005; October 17 - November 22, 2006; October 10 - December 10, 2012; September 26 - November 28, 2014.  In the 2012 and 2014 observations we filled unused fibers with WISE sources (Wright et al. 2010). To obtain a highly complete survey over the entire F1 field we applied the Roll et al. (1998) observation planning software and we revisited each Hectospec positioning more than 7 times on average. 

\begin{figure}
\centerline{\includegraphics[width=7.0in]{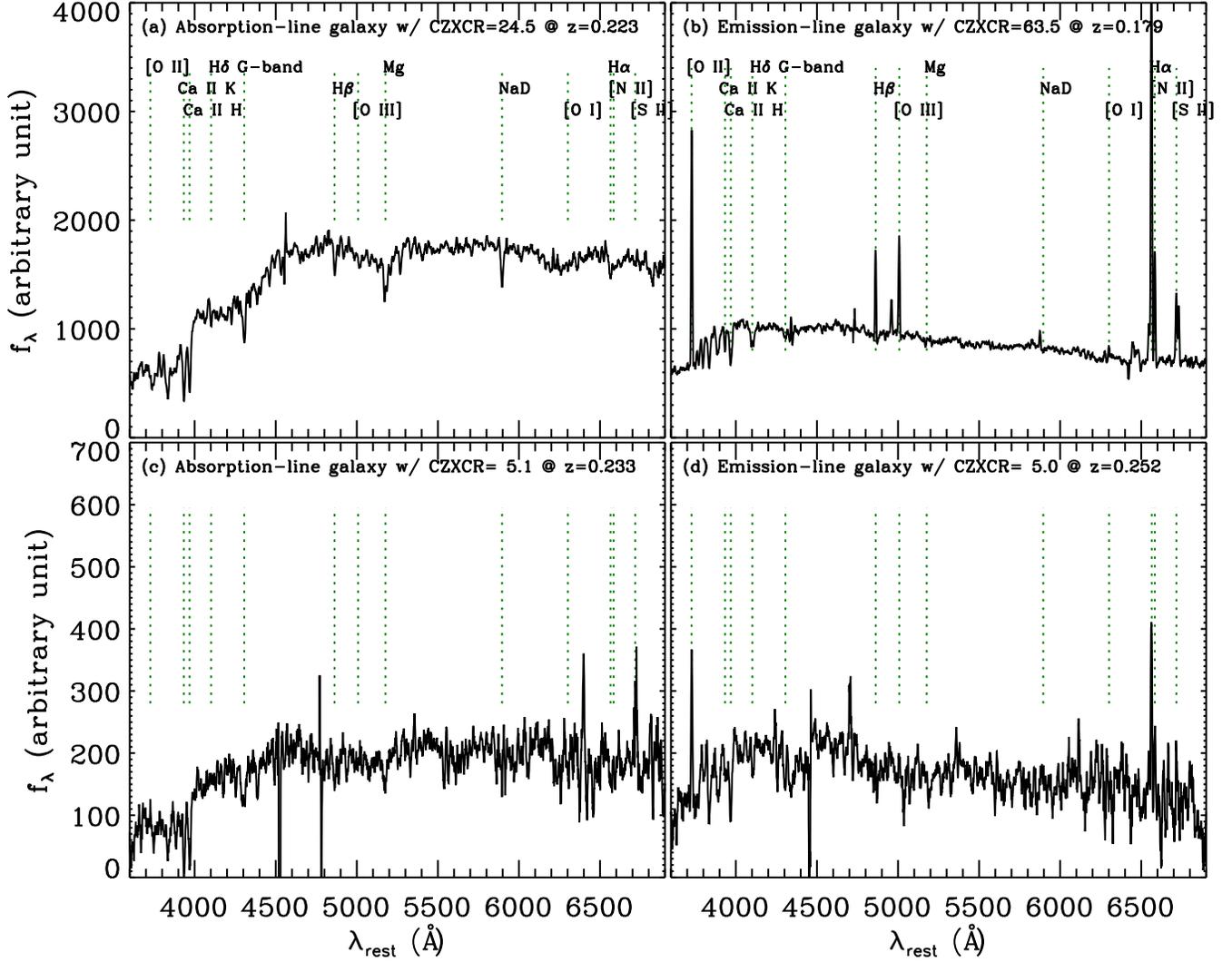}}
\vskip 5ex
\caption{Sample absorption-line (left) and emission-line spectra (right) demonstrating the 
range of quality (cross-correlation coefficient) at $z \sim 0.22$. Labels indicate 
major spectral features; unlabeled spikes are
badly subtracted night sky lines.  
\label{Fegsp}}
\end{figure}

We use the same oberving protocols and reduction procedures for the F1 and F2 fields. Figure \ref{Fegsp} shows typical high and low signal-to-noise spectra acquired in our 0.75 to 2 hour integration time. We show the spectra for a window in the rest frame. The wavelength range covered by Hectospec in the observer's frame is  3,700 --- 9,100 \AA\ with a resolution of {$\sim$5 \AA}. The Hectospec fibers have a 1.5$^{\prime\prime}$ diameter.

We reduced the 2005, 2006, and 2012 data with the Mink et al. (2007) Hectospec pipeline and derived redshifts with RVSAO (Kurtz \& Mink 1998; see also Fabricant et al.2005). We reduced the 2014 observations
with HSRED v2.0 developed by the SAO Telescope Data Center. This pipeline is a
revision of the IDL pipeline originally written by Richard Cool (see
http://www.mmto.org/book/export/html/55). There is no difference between the redshifts derived from the two pipelines.

In the analysis of the F2 data, we used repeat Hectospec observations of 1651 pairs of spectra of 1651 unique objects to compute internal errors in the redshift. We used the generally brighter ovelapping SDSS galaxies to estimate an external error. For emission line objects the internal error (normalized by (1 + z)) is 48 km s$^{-1}$ and for absorption line objects it is 24 km s$^{-1}$. The typical external error,  which may be underestimated from  comparisons with brighter SDSS objects, is 37 km s$^{-1}$ regardless of spectral type.  There is a small offset between the SDSS and MMT redshifts,
$\Delta{z}/(1 + z)$ = 3.4 $\pm$ 3.9 km s$^{-1}$. 
These errors also apply for F1; the instrument, the procedures, and the reduction are
essentially identical.

\begin{figure}
\centerline{\includegraphics[width=7.0in]{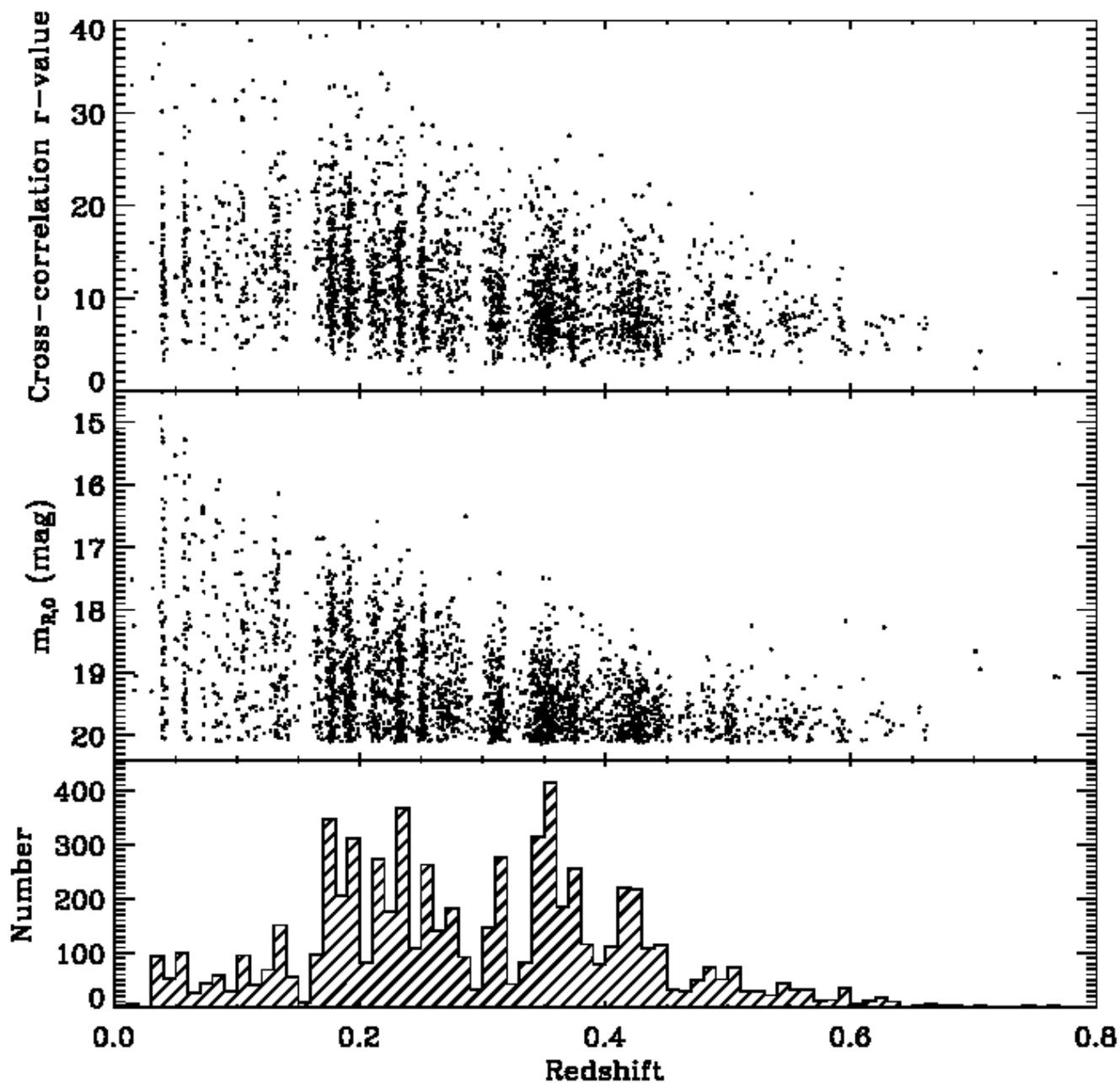}}
\vskip 5ex
\caption{Cross-correlation $r$-value (Tonry \& Davis 1979), a redshift  quality indicator, as a function of redshift (upper panel). The center panel shows apparent R$_0$ magnitude as a function of redshift. We display only 50\% 0f the data for clarity. The lower panel shows a redshift histogram in bins of 
$\Delta{z} = 0.01$. 
\label{BFzxcr}}
\end{figure}

As a result, in part, of variable conditions, the quality of spectra that yield acceptable redshifts varies significantly.  
The pipeline provides a standard measure of the quality of the spectrum, $r_{TD}$, a measure of the width of the cross-correlation peak originally defined by Tonry \& Davis (1979). Based on the value of $r_{TD}$ we compute the redshift error as in Kurtz \& Mink (1998). Figure \ref{BFzxcr} shows  $r_{TD}$ as a function of 
the extinction corrected R$_0$ apparent magnitude for the F1 field.  Kurtz \& Mink
(1988) show (their Figure 8) that $r_{TD}$ measures the signal-to-noise of the spectrum. For
surveys like F1 that span a significant redshift range $r_{TD}$ is a better indicator of the quality of the redshift than signal-to-noise at a fixed wavelength.  This Figure is in the same format as Figure 2 in Geller et al. (2014) for F2 to facilitate direct comparison. Here we limit the figure to the highly complete R$_0$ $\leq 20.2$ sample in F1.

Although Figure \ref{BFzxcr} suggests that we could select reliable redshifts based merely on the value of $r_{TD}$, we inspect each spectrum visually after the pipeline processing. We then
classify the spectra with three flags: Q for high quality, ? for dubious cases, and X for completely unacceptable data. We report only Q redshifts here just as we did for F2. 
Visual inspection identifies (1) spectra corrupted by badly subtracted night sky, (2) cases where there are two objects at different redshift in a single fiber, and (3) a small number of quasars. Otherwise the visual classification is essentially a reflection of 
$r_{TD}$. 

In the following sections we limit discussion to the redshift survey for 
R$_0$$ \leq 20.2$, the faintest limiting magnitude where the completeness in F1 significantly exceeds 90\%. In Table \ref{tab20.2} and Table \ref{tabfaint} we list all of the redshifts we measured in the F1 field. In total, the Tables include 9861 redshifts; 9639 are new measurements with Hectospec, 185 are from the SDSS DR12 (Alam et al. 2015),
and 37 are from NED.  Table \ref{tab20.2} is the magnitude limited sample we analyze and Table \ref{tabfaint} contains redshifts we measured for fainter objects. The Tables include the SHELS ID, the SDSS ObjID, the total extinction corrected R$_0$ magnitude from the DLS along with its error, the redshift and its formal error derived from the $r_{TD}$ value. The Table includes a flag  if
the R-magnitude is converted from an SDSS r-band magnitude or if the redshift comes from the SDSS or NED. We also indicate whether the source is classified as a point source in the SDSS 
based on the probPSF parameter in the SDSS database.  These SDSS point sources were  observed as WISE QSO candidates to fill unused fibers. We do not use any of these point sources in the analysis below and include them solely to publish the redshifts.

As for the F2 field, Tables \ref{tab20.2} and \ref{tabfaint} also include three derived quantities: D$_n$4000, the stellar mass and its error, and the metallicity (for emission-line galaxies in the redshift
range 0.2$ < z < 0.38$ (as we discuss in Section 3.1 below). Section 2.3 of Geller at al. (2014) describes the
derivation of these quantities and we do not repeat the discussion here. Our procedures for F1 are identical to those for F2. The computation of D$_n$4000
is based on the procedures described in Fabricant et al. (2008) following the definition of Balogh et al. (1999). We derive stellar masses according to the procedures described by Zahid et al. (2013) based on the
Le Phare code written by Arnouts \& Ilbert (Arnouts et al. 1999; Ilbert et al. 2006). The derivation of metallicities follows Zahid et al. (2013, 2014) based on the R23 line ratio calibration by Kobulnicky \& Kewley (2004). Briefly, to construct the sample of star-forming galaxies we require a S/N$>$3 in the line flux measurements of [OII] $\lambda$3727, 3729, H$\beta$, H$\alpha$, and [NII] $\lambda$6584. We remove AGN from the sample of star-forming galaxies based on the BPT method (Baldwin et al. 1981, BPT) as updated by Kewley et al. (2006).

Again as in the F2 field we do not report unphysical values of D$_n$4000 (D$_n$4000 $<0$ or D$_n$4000$ >$3. These values result from poor spectra that are merely adequate to yield a redshift. In Section \ref{distdn4000} we compare the distributions of D$_n$4000 as a function of
stellar mass and redshift for the F1 and F2 fields. The fraction  of objects with unmeasurable D$_n$4000 in both fields
(as detailed in Section \ref{distdn4000}) is so small that the effect on this comparison is negligible.

To make the definition of the sample with R$_{0} \leq 20.2$ clear, we list the galaxy candidates without redshifts to this limit in Table \ref{tabnoz}. Among these candidates, 
visual inspection of  the DLS images suggests that $\lesssim 4$\% of these remaining objects could be stars. In Table \ref{tabstar}, for completeness, we list the objects
that we identify spectroscopically as stars.

\subsection {Redshift Survey Completeness}
\label{completeness}
For both the F1 and F2 fields, Table \ref{numbers} lists the number of photometric objects and the number of spectroscopically confirmed galaxies for several magnitude limited samples. We focus on comparing two samples with high spectroscopic completeness to the extinction corrected R$_0$ = 20.2. In fact, the completeness listed in Table \ref{numbers} is a lower limit because 3-4\% of the objects without a measured redshift are stars.  We also list the number of redshifts (published here) for galaxies fainter than our initial nominal limit R = 20.6.

At face value there are several interesting aspects of the numbers in Table \ref{numbers}.   First, at every magnitude limit, the raw number of galaxies per unit area is
larger for F2 than for F1. As the extinction corrected sample shows, this difference is not a result of the greater extinction for F1. We show in Section \ref{variance} that this difference is actually consistent with the expected cosmic variance.

\begin{figure}
\centerline{\includegraphics[width=7.0in]{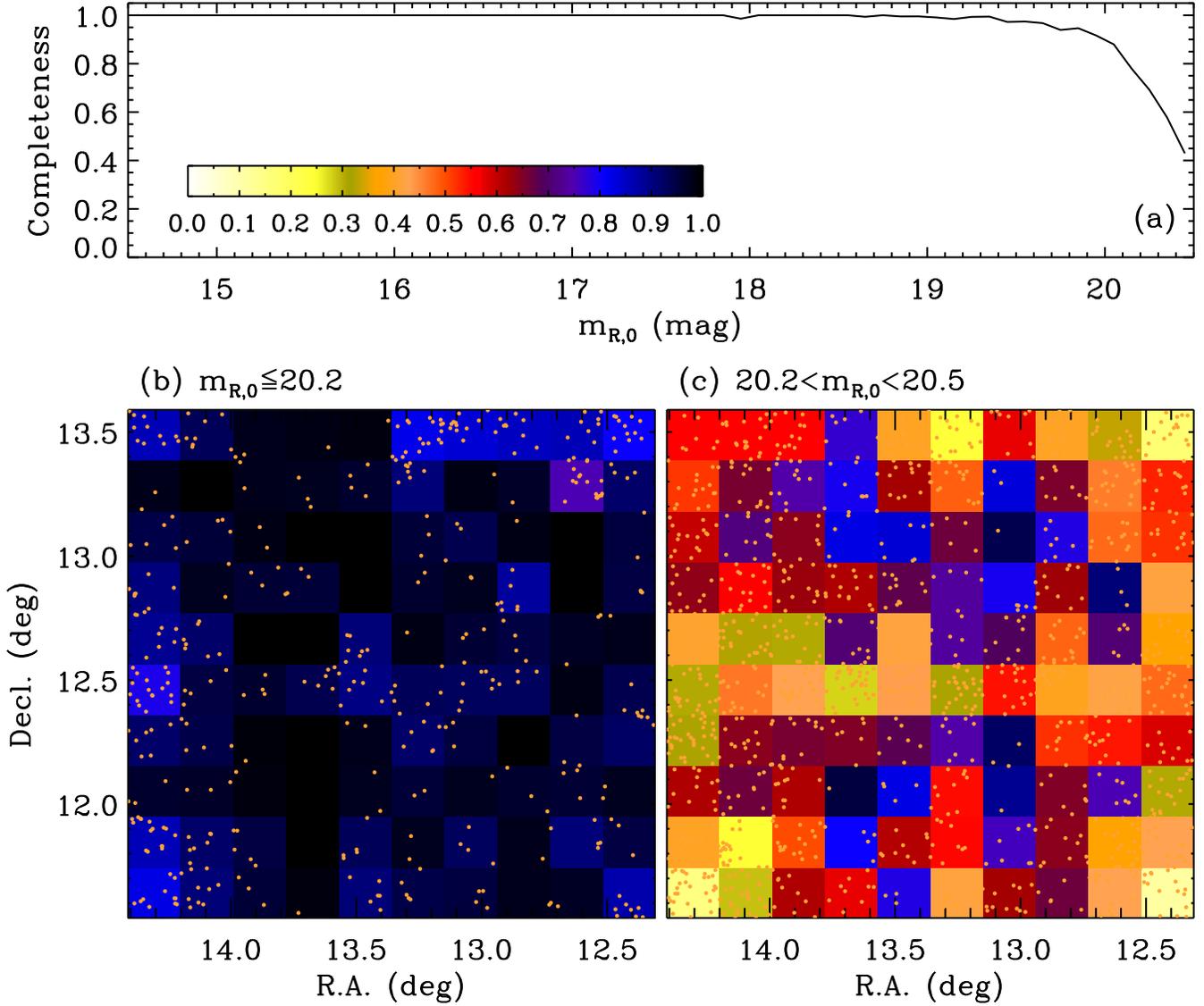}}
\vskip 5ex
\caption{Completeness of the SHELS redshift survey of the DLS F1 field. The upper panel shows the completeness as a function of DLS extinction corrected R$_0$ magnitude. The color bar shows the completeness fractions for the spatial completeness displays in the lower two panels. The lower left panel shows the completeness in 12$\times$12 arcminute bins for 
galaxies with R$_0 \leq 20.2$. The yellow points indicate galaxies in the photometric sample without a measured redshift. The right hand plot shows the  completeness in the interval 20.2$ <$R$_0$$<20.5$ in the same format. Note that for the R$_0 \leq 20.2$ survey the most significant incompleteness occurs at the
corners and edges of the field.
\label{BFcomplete}}
\end{figure}

Figure \ref{BFcomplete} provides a more detailed picture of the completeness of the F1 field. The Figure is in the same format as Figure 4 of Geller et al. (2014); corrected for extinction, the magnitude limits in the corresponding panels are essentially the same. 
The 422 objects without a redshift in F1 are clearly not uniformly distributed over the field. In the map for the survey limited to R$_0$ = 20.2 (left panel) most of the pixels that are $\lesssim 90$\% complete lie along the edges, but there are a few within the central 8 $\times$ 8 pixel region. The top panel shows the steep drop in average completeness as a function of limiting extinction corrected magnitude. The right-hand panel shows the highly variable, relatively poor completeness of the F1 field in the interval 
$20.2 < $R$_0$ $< 20.5$. This panel substantiates our decision to limit discussion of this
region to the brighter magnitude limit.

\begin{figure}
\centerline{\includegraphics[width=6.5in]{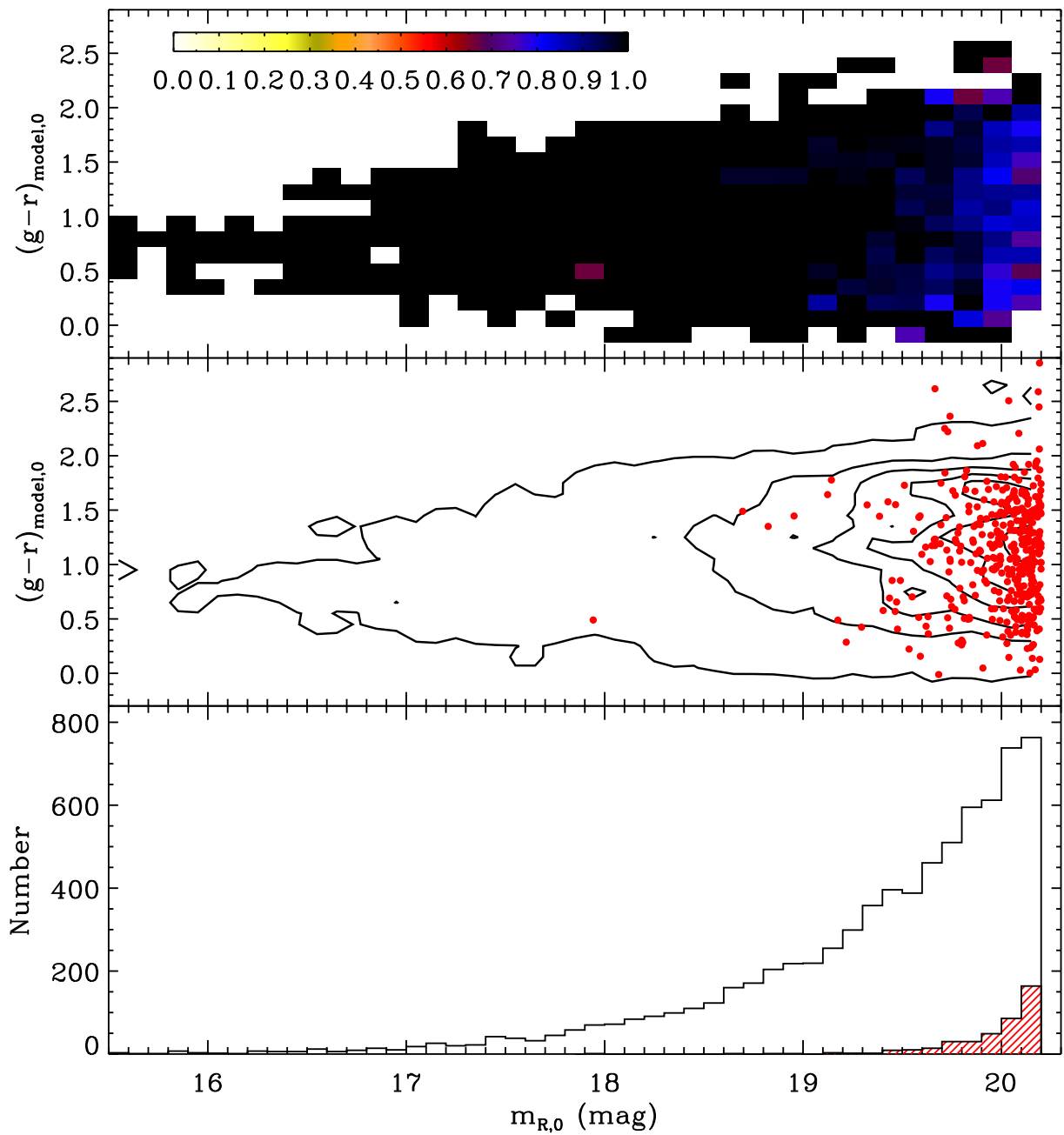}}
\vskip 5ex
\caption{Completeness of F1 as a function of color and R$_0$ magnitude (upper panel). Color-magnitude diagram for the 422 objects without a redshift (middle panel) 
with R$_0 \leq 20.2$. Contours indicate the relative density of objects with a redshift; the absence of a slope as a function of R$_0$ suggests that there is little obvious color bias in these objects. The bottom panel shows the number of objects with redshifts in the SHELS survey (open histogram) and the number of unobserved galaxy candidates (red hashed histogram) as a function of the R$_0$ magnitude.
\label{Fcmr}}
\end{figure}

The color-magnitude diagram (Figure \ref{Fcmr}) for the objects without a redshift shows that at this bright limit, there is essentially no color dependence among the missing objects (middle panel).  As expected, the number of objects without spectroscopy increases significantly for the faintest objects
in the survey reaching nearly 25\% in the faintest 0.1 magnitude bin (top and bottom panels). Figure 5 of Geller
et al. (2014) shows the analogous plot for F2 where only 273 objects with R$_0$$\leq 20.2$ lack a redshift; the salient features (the lack of color dependence and the rise in incompleteness with apparent magnitude) are similar.

\begin{figure}
\centerline{\includegraphics[width=6.5in]{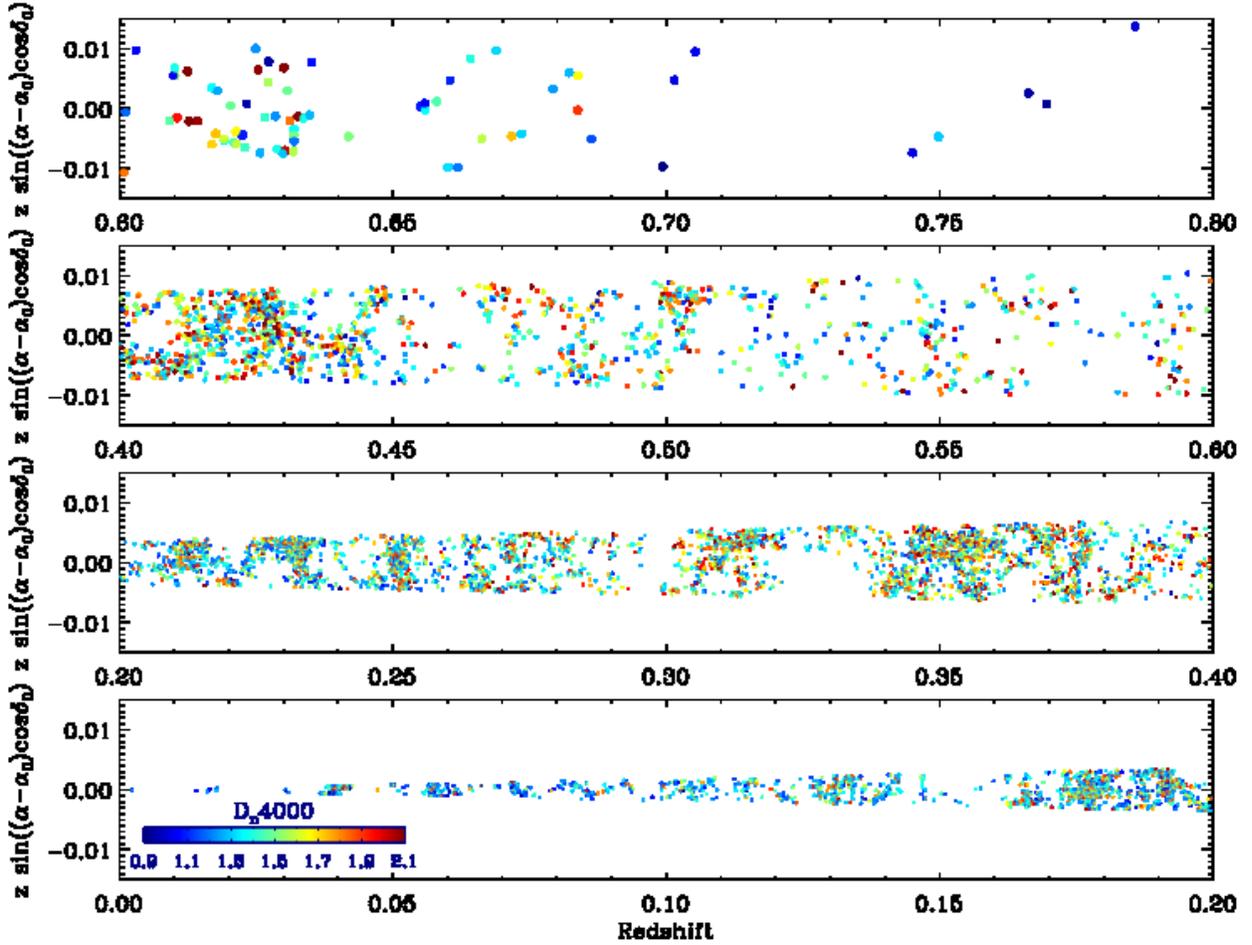}}
\vskip 5ex
\caption{Cone diagram for the R$_0\leq$ 20.2 SHELS F1 survey projected on the R.A.$_{2000}$ direction. The color coding indicates the value of D$_n$4000. In the low density regions, galaxies with D$_n$4000 $\lesssim$ 1.5 predominate as expected. The online journal includes a video display of the data. The color-coding of the video is in broader bins: D$_n$4000 $< 1.3$ (blue), 1.3 $\leq$ D$_n$4000 $< 1.7$ (green), and D$_n$4000 $\geq$ 1.7 (red). 
\label{BraFcone}}
\end{figure}

The cone diagram of F1 projected in R.A. (Figure \ref{BraFcone}) reveals the characteristic  cosmic web
structure (Geller \& Huchra 1989). The color coding as a function of D$_n$4000 is the same as in the cone diagram for F2 in Figure 6 of Geller et al. (2014). In both fields, the tendency toward galaxies with younger populations in lower density regions is evident in the bluer color of the points. In F1 there are no prominent concentrations of rich clusters; in F2, the A781 complex dominates the survey at z$\simeq$ 0.29-0.30. A finger corresponding to a cluster
at z$\sim 0.35$ is visible in the cone diagram. A movie with broader binning in 
D$_n$4000 shows the 3D structure in F1 and highlights the $z = 0.35$ cluster by zooming in on it. Section \ref{variance} contains a more extensive discussion of clusters in the two fields.

\section {Comparing the F1 and F2 Fields of the DLS}
\label{comparison}
Although the construction of the surveys of F1 and F2 is not identical, the surveys are sufficiently similar to provide an interesting comparison of two widely separated similarly observed regions 
of the universe. Our goal is to investigate a few of the differences and similarities between the fields to highlight the quality of the data and to provide a benchmark for
further scientific applications. These applications include but are not limited to planning, calibration, and analysis of color-selected surveys (see e.g. Damjanov et al. 2015; Geller \& Hwang 2015).

\begin{figure}
\centerline{\includegraphics[width=7.0in]{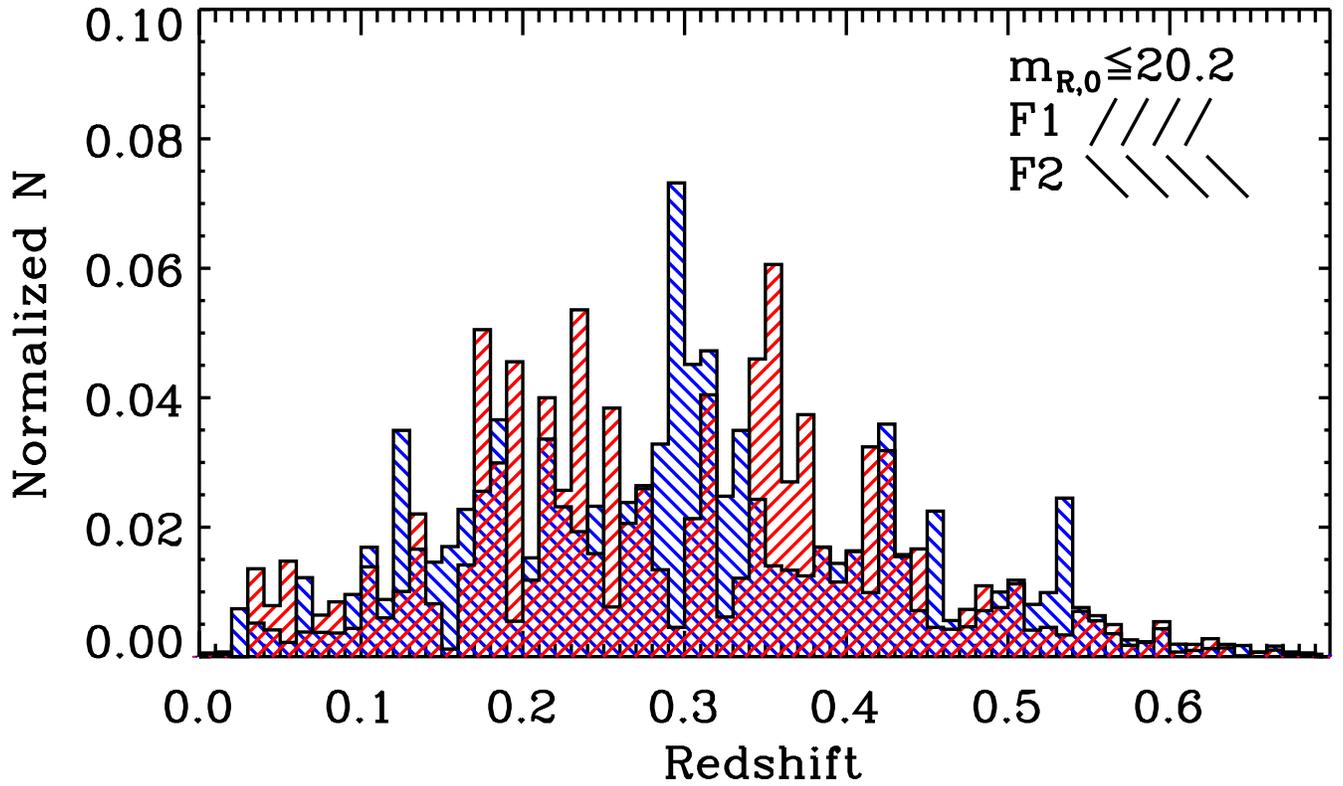}}
\vskip 5ex
\caption{Comparison of normalized redshift histograms for the F1 (red) and F2 (blue) fields of the DLS.
The surveys are both limited to R$_0$ = 20.2. Bins are $\Delta{z} = 0.01$. Note the marked differences in the histograms, particularly in the range $0.25 < z < 0.4$.
\label{Bnzf1f2}}
\end{figure}

Two straightforward figures provide an introduction to the salient differences between the
two fields. First, Figure  \ref{Bnzf1f2} shows the normalized redshift histograms for the two regions. The difference is striking. The F2 field (blue histogram) has a broad, prominent peak
centered near $z \sim 0.3$. The most prominent peak in the F1 field (red) is near $z \sim 0.35$.  A K–S test rejects the hypothesis that the distributions of the two samples are extracted from the same parent population at a confidence level of 99.9\%.

\begin{figure}
\centering
%\begin{tabular}{lr}
%\begin{tabular}{cc}
%\includegraphics[width=4.5in]{Subtest.ps}
%\includegraphics[width=5.5in]{plotStruct3.eps}
\begin{tabular}{cc}
\includegraphics[width=3.2in]{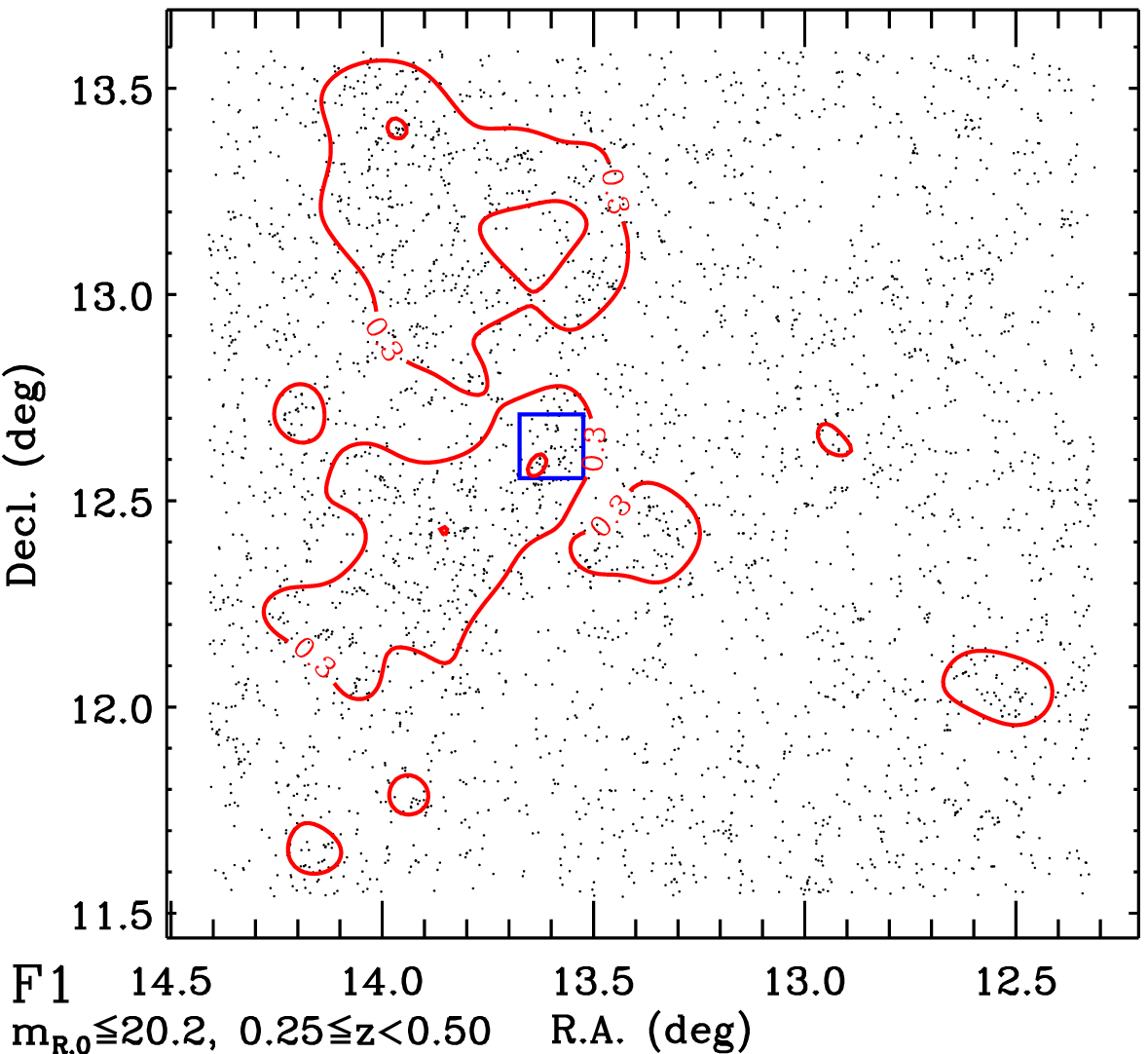}
\includegraphics[width=3.2in]{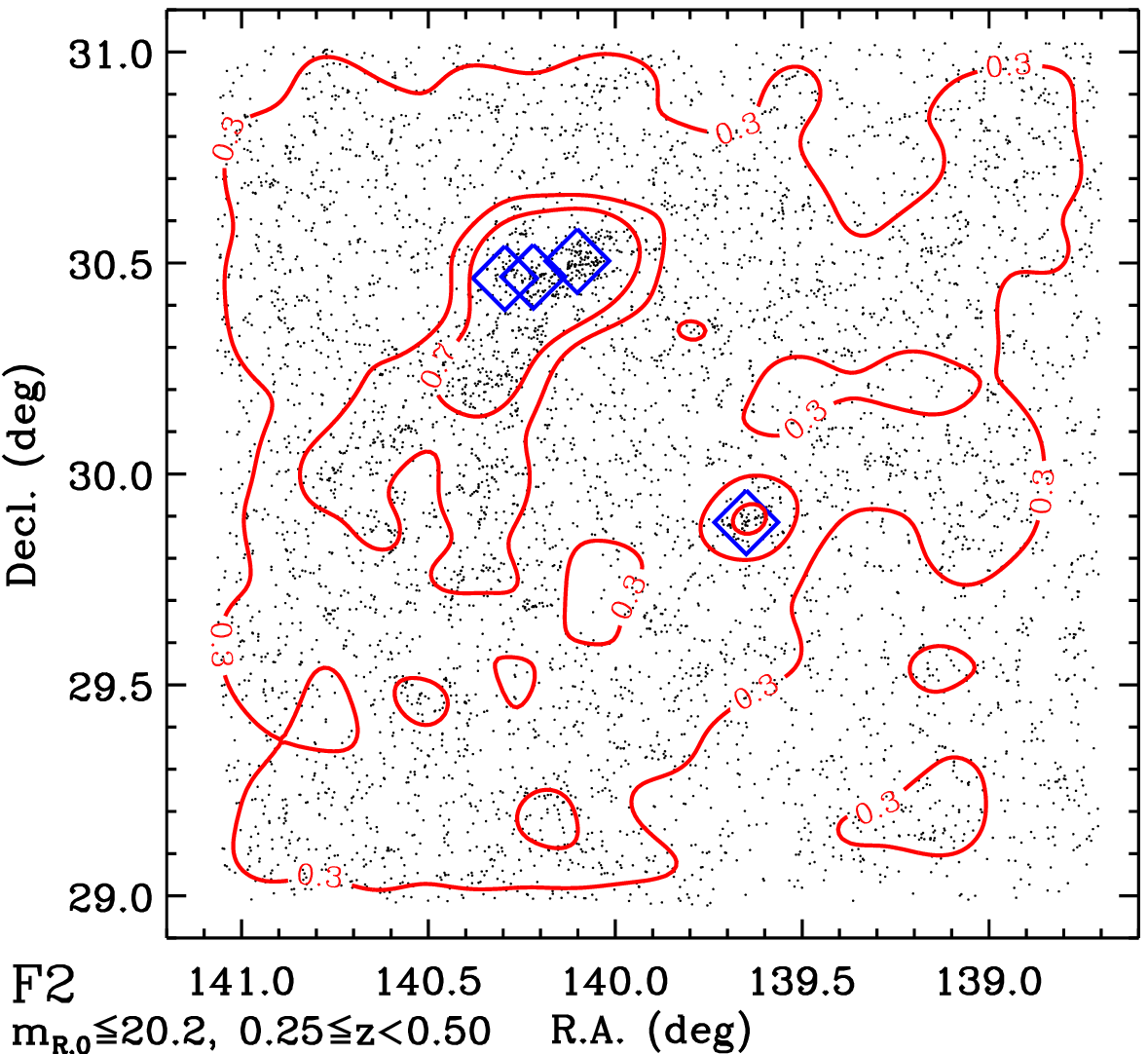}
\end{tabular}
%\end{tabular}
\caption{Distribution on the sky of galaxies with $0.25 \leq z < 0.5$ in the F1 (left) and F2 (right) redshift surveys. Each of the black points represents a galaxy with R$_0 \leq 20.2$. The isodensity
contours highlight the striking difference between the two fields: for F1 the contours are
0.3 and 0.5 gals arcmin$^{-2}$ and for F2 they are 0.3, 0.5, and 0.7 gals arcmin$^{-2}$. Black Diamonds in F2 indicate the four clusters with $0.25 \leq z < 0.5$ that are cleanly detected by three methods:
x-ray, weak lensing, and the SHELS survey. All of these systems lie in regions of the  highest galaxy surface number density. In the same redshift range in F1 there is a single cluster
candidate at $ z = 0.35$ marked by a box  and identified only by galaxy counts and spectroscopic data.}
%\end{sidewaysfigure}
\label{maps}
\end{figure}

Figure \ref{maps} shows the distribution of galaxies with $0.25 \leq z < 0.5$ on the sky for both
F1 and F2. The galaxy isodensity contours show that F1 lacks any regions that reach the highest projected number density  in F2. For F2 the map shows the four high confidence clusters (diamonds)
detected by three methods: SHELS spectroscopy, x-ray and weak lensing. All of these systems lie in the highest density regions of the map. In the F1 field we 
show one candidate cluster that is indicated only by SHELS and galaxy counts. We discuss the
clusters further in Section \ref{clusters}.

\subsection {The Mass-Metallicity Relation}
\label{mzrelation}
The combination of spectral properties with stellar masses of galaxies provides a powerful basis for understanding the nature of galaxies and their evolution (e.g. Brinchmann \& Ellis 2000; Kauffmann et al. 2003; Bell et al. 2003). Brinchman \& Ellis (2000) emphasize the power of exploring galaxy evolution with cosmic time by combining a more transitory spectral signature with the
generally slowly varying stellar mass derived from multi-band imaging.

Here we combine stellar masses derived from $ugriz$ imaging with the results of strong-line metallicity diagnostics to explore the robustness of the mass-metallicity (MZ) relation for survey regions
like the DLS fields. In Section \ref{distdn4000} we examine the distribution of D$_n$4000 as a function of stellar mass and redshift as another measure of 
variations among fields similar to those probed by the DLS.

As in previous metallicity analyses of these fields (Zahid et al. 2013; Geller et al. 2014; Zahid et al. 2014), we compute metallicities for star-forming galaxies in the redshift range 0.2$ < z < 0.38$.  These limits are set by the 
1.5$^{\prime\prime}$ fiber size and by the bandpass  of Hectospec. The fiber aperture is too small to include enough of the galaxy light ($\gtrsim 20$\%) for z$\leq$ 0.2 (see Kewley et al 2005), and H$\alpha$ shifts out of the bandpass for redshifts $\gtrsim$ 0.38 making the exclusion of AGN difficult. 
As for the F2 field (Geller et al. 2014), we include metallicities for individual objects in Table \ref{tab20.2}.  

\begin{figure}
\centerline{\includegraphics[width=7.0in]{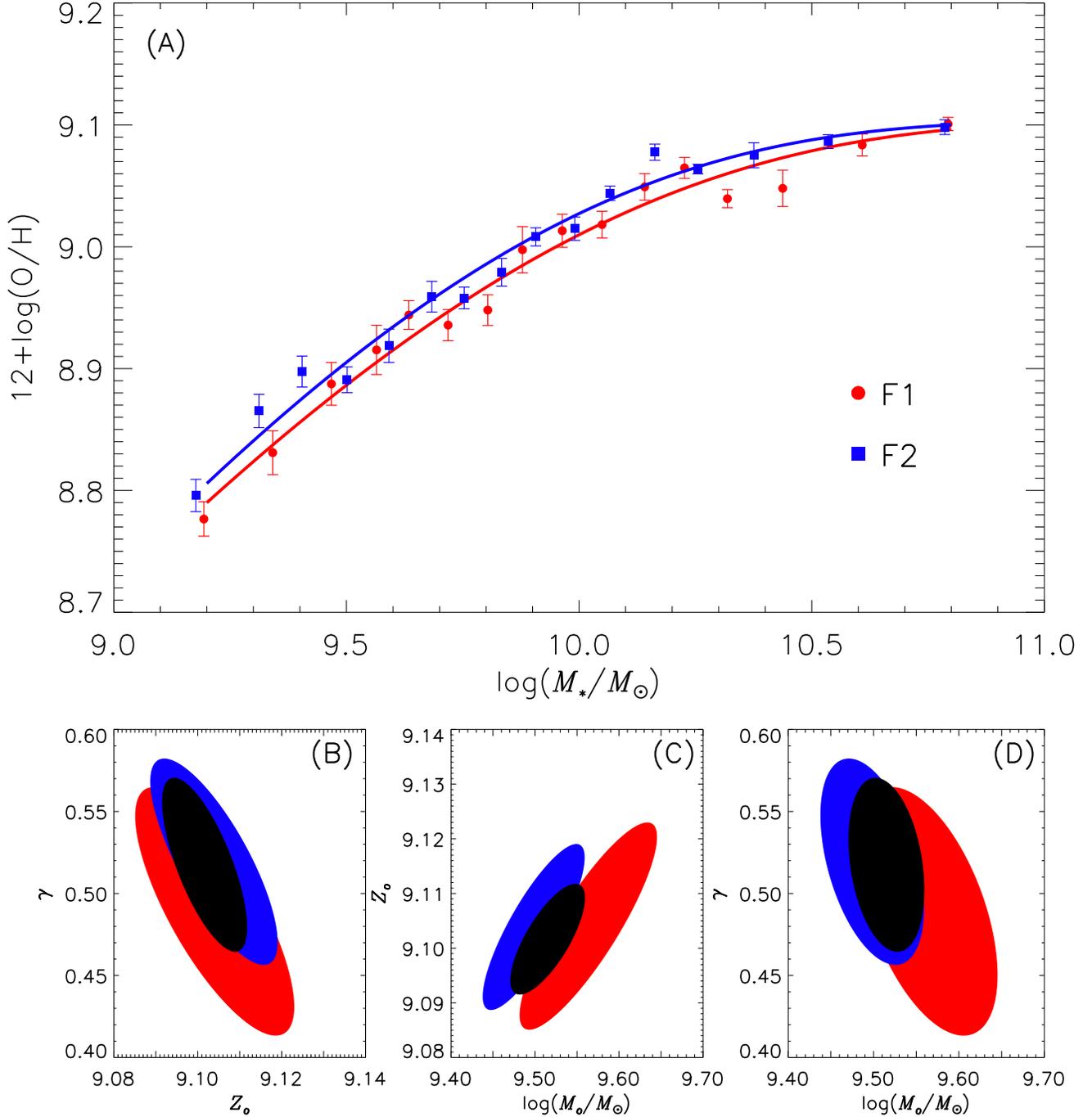}}
\vskip 5ex
\caption{Mass-metallicity relations (upper panel) for independent samples
from F1 (red) and F2 (blue).  The points show the median metallicity in each mass bin and the error bars are the bootstrapped 68\% confidence limits. The curves are fits of equation (1). The lower panel shows the 95\% confidence error ellipses for combinations of
the three parameters of the fit: Z$_0$, $\gamma$, and M$_0$/M$_\odot$. The black error ellipse shows the 95\% confidence limits for the combined F1 plus F2 sample analyzed by Zahid et al. (2014).
\label{Fmzrel}}
\end{figure}

Figure \ref{Fmzrel} shows the MZ relations for the two fields. The curves are fits of the form
\begin{equation}
12 + log(O/H) = Z_0 + log \bigl[1-exp\bigl(-\bigl[{M_*\over{M_0}}\bigr]^\gamma\bigr)\bigr] 
\end{equation}
\noindent Zahid et al. (2014) discuss the physical interpretation of this form for the MZ relation. We simply note that $Z_0$ is the saturation metallicity. For stellar masses $M_* \gtrsim M_0$, the metallicity approaches the saturation limit, $Z_0$. The slope, $\gamma$ characterizes the MZ relation for M$_* <<$M$_0$. 

Table \ref{tblMZ} lists the parameters of the MZ relation fits in Figure \ref{Fmzrel}. The lower panel of Figure \ref{Fmzrel} shows the 95\% confidence error ellipses for three samples: F1 (red), F2 (blue), and the F1 plus F2 sample from Zahid et al. (2014) (black). We quote values of the fit parameters along with the formal 1$\sigma$ errors in the Table.  We also include the combined result from Zahid et al. (2014) in Table \ref{tblMZ} for reference.
 
The results for F1 and F2 are two independent measurements of the MZ relation in the redshift range 0.2$ < z < 0.38$. They provide some assessment of the potential impact of cosmic variance on the determination of the MZ relation for samples covering a volume of $\sim$ 10$^6$ Mpc$^3$ in each field. The difference between the MZ relations for the two fields is small.

All of the  95\% confidence ellipses for the MZ relation parameters overlap
(Figure \ref{Fmzrel}, lower panel). The parameter most sensitive to evolutionary effects is the transition stellar mass, M$_0$ (Zahid et al. 2014). Thus it is perhaps not surprising that the Z$_0$ vs log(M$_0$/M$_\odot$) error ellipses have the smallest fractional overlap, but the offset is only at the $\lesssim\ $1.6$\sigma$ level.  

Peng \& Maiolino (2014) use the SDSS to show that the MZ relation has some environmental dependence: satellite (generally less massive) galaxies have greater metallicity in denser environments.   This trend is generally consistent with the offset in the  Z$_0$ vs log(M$_0$/M$_\odot$) error ellipses and the offset in the corresponding MZ relations of Figure \ref{Fmzrel}. 

The lower panel of Figure \ref{Fmzrel}  also shows confidence ellipses for the F1 plus F2 sample which effectively averages over the differing overall densities of the F1 and F2 fields. The error ellipses for the combined sample are located, mainly and not surprisingly, where the ellipses for F1 and F2 overlap. The corresponding parameters for the F1 plus F2 sample listed in Table \ref{tblMZ} are the best estimate of the MZ relation for the entire SHELS survey covering the redshift range
0.2 $ < z < 0.38$. The small difference between the independent estimates based on F1 and F2 suggests that the result  for the combined sample is a robust representation of the properties of star-forming galaxies  in this redshift range.

\subsection{Distributions of D$_n$4000}
\label{distdn4000}
The spectral indicator D$_n$4000 has a rich history as a measure of galaxy properties and their evolution (Mignoli et al. 2005; Bundy et al. 2006; Roseboom et al. 2006; Noeske et al. 2007; Vergani et al. 2008; Freedman Woods et al. 2010; Moresco et al. 2010; Moustakas et al. 2013; Moresco et al. 2013; Geller et al. 2014). Here we examine the distributions of D$_n$4000 as a function of both stellar mass and redshift for the F1 and F2 fields. As for the MZ relation, we explore the impact of the different overall galaxy density in the two fields on these distributions.
We seek to link any differences with differences in features of the large-scale galaxy distribution in
the two fields. 

\begin{figure}
\centerline{\includegraphics[width=6.0in]{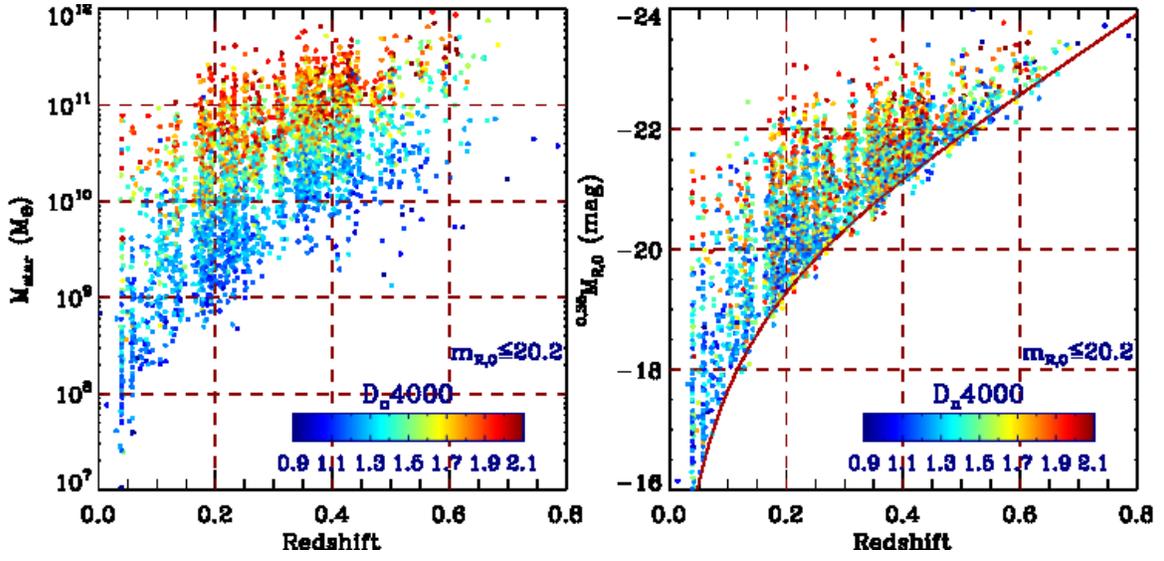}}
\vskip 5ex
\caption{Stellar mass as a function of redshift (left) and K-corrected (to $ z = 0.35$) R$_0$ absolute magnitude as a function of redshift (right) for the survey limited to R$_0$ = 20.2. In  both panels galaxies are color-coded by D$_n$4000. We display only 50\% of the data for clarity.   
\label{Fmassz}}
%\end{sidewaysfigure}
\end{figure}

Figure \ref{Fmassz} (left) shows stellar mass as a function of redshift for F1. The points representing galaxies in the survey are color-coded by D$_n$4000. The segregation of D$_n$4000 with stellar mass is obvious; galaxies with larger stellar mass have larger D$_n$4000 suggesting older ages and perhaps higher metallicities. There is also the known evolutionary trend 
that large values of D$_n$4000 occur for galaxies with lower  stellar mass at
lower redshift. The right-hand panel of Figure \ref{Fmassz} shows the K-corrected absolute magnitude (shifted to $ z = 0.35$) for the F1 survey. Again the points are color-coded with the value of D$_n$4000. The contrast between the left and right panels demonstrates the known advantage of using stellar mass rather than absolute magnitude to characterize galaxy populations. 

The advantages of D$_n$4000 include the strength of the feature, its essential redshift independence (in contrast with colors), and its insensitivity to reddening. 
The simplest interpretation  of the D$_n$4000 indicator  as a population age indicator is complicated somewhat by metallicity dependence (e.g. Balogh et al. 1999; Kauffmann et al 2003). 

Based on repeat Hectospec observations, Fabricant et al. (2008) show that the error in the
D$_n$4000  is 0.045 times the value of the index. Comparisons with measurements derived from the larger SDSS fiber apertures show that there is no apparent bias in the Hectospec 1.5$^{\prime\prime}$-fiber D$_n$4000 values relative to the 
3$^{\prime\prime}$-fiber SDSS values. Although there may be subtle aperture effects as a function of redshift and stellar mass, we assume the values we derive for the galaxies in F1 and F2 are representative. Our goal here is comparison of the distributions of the indicator in the two fields where we segregate objects by both redshift and stellar mass; thus aperture effects should be irrelevant.

\begin{figure}
\centerline{\includegraphics[width=7.0in]{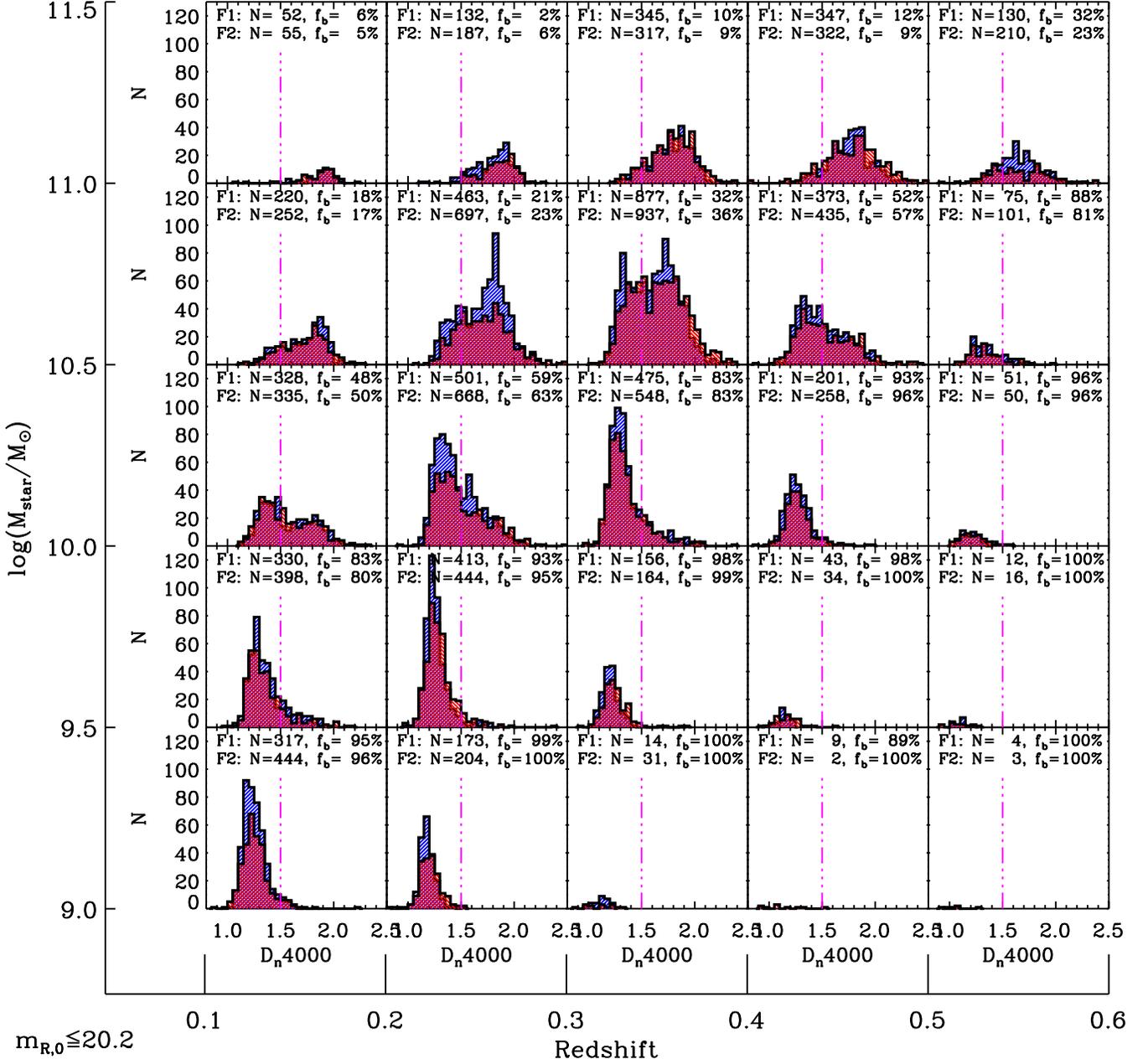}}
\vskip 5ex
\caption{Histograms of D$_n$4000 in bins of stellar mass and redshift for the F1 (red) and F2 (blue) fields.  At fixed stellar mass, the expected evolutionary effects appear in both fields; the fraction of low D$_n$4000 (probable star-forming galaxies) increases with redshift at fixed stellar mass. Remarkably the fractions of galaxies with D$_n$4000 $ < 1.5$ ($f_b$) are consistent to within the errors in nearly all bins. Note, however, the excess absolute counts in the F2 field in the 0.2$ < z < 0.3$ bin. This bin contains the main concentration of the A781 complex.
\label{subdn4000}}
\end{figure}

As we noted in Section \ref{spectroscopy}, the spectra of some objects do not provide a measure of D$_n$4000. Over the range in redshift and stellar mass explored in Figure \ref{subdn4000}, there are only 3 objects in the F1 field (0.1\% of the total survey) without a measure of D$_n$4000. In F2 there are 198 objects (2.7\% of the sample) without a measure of D$_n$4000. Only the bin 0.5 $ < z < $ 0.6, 11$ < log(M_{star}/M_\odot) <$ 11.5  is significantly affected by this incompleteness: 11\% of the galaxies in this
bin lack a secure D$_n$4000. For all other bins, the
small fractions  of missing objects do not affect the comparison of D$_n$4000 distributions.

Figure \ref{subdn4000} shows the D$_n$4000 distributions as a function of redshift and stellar mass for F1 (red) and F2 (blue). The histograms are not normalized to emphasize the impact of large-scale structure on the occupancy of these bins. In each panel we list the number of galaxies in each survey and the fraction, $f_b$, with D$_n$4000$< 1.5$. The fiducial value 1.5 effectively separates star-forming from quiescent galaxies (Freedman Woods et al. 2010). 

Remarkably, the values of $f_b$ are consistent for the two samples in nearly every redshift-mass bin in Figure \ref{subdn4000} even though the numbers of galaxies in the bin may sometimes differ significantly. Well-populated bins with the most significant differences in raw counts
occur in the ranges 0.2 $ < z <$ 0.3,  10$ < log(M_{star}/M_{\odot}) <$ 11.5
and 0.1$ < z < 0.2$, 9$ < log(M_{star}/M_{\odot}) <$ 9.5. The presence of massive clusters of galaxies (see Section \ref{clusters}) accounts for the difference at 0.2$ < z < 0.3$; in F2 the
A781 complex of two massive clusters contributes substantially in this range; there is no comparable system in F1. In the range 0.3$ < z < 0.4$, the A781 complex still contributes in
F2 and F1 contains a massive system with mean redshift 0.35 (see Section 3.4). At 0.2$ < z < 0.3$, the
most significant differences occur in the stellar mass range 10$ < log(M_{star}/M_\odot) <$ 11.5 as a result of the stellar mass dependence of galaxy clustering (e.g. Bielby et al. 2014); galaxies of greater stellar mass preferentially inhabit denser regions.
In the low redshift bin, 0.1 $ < z < 0.2$, the greater absolute abundance of galaxies with 
9$ < log(M_{star}/M_\odot) <$ 9.5 in F2 reflects the presence of a  z$ \sim 0.125$ structure where the survey is deep enough to include these low stellar mass galaxies. It is worth emphasizing that the
DLS field selection avoids regions with clusters at $z \lesssim$ 0.1, but the presence/absence of structures near this limit affects the galaxy count particularly at low stellar mass. The largest differences in the relative abundances of galaxies
in the bins of Figure \ref{subdn4000} primarily reflect the differences in the populations of clusters near the peak sensitivity of the redshift survey. 

\begin{figure}
\centering
\begin{tabular}{lr}
\includegraphics[width=3.2in]{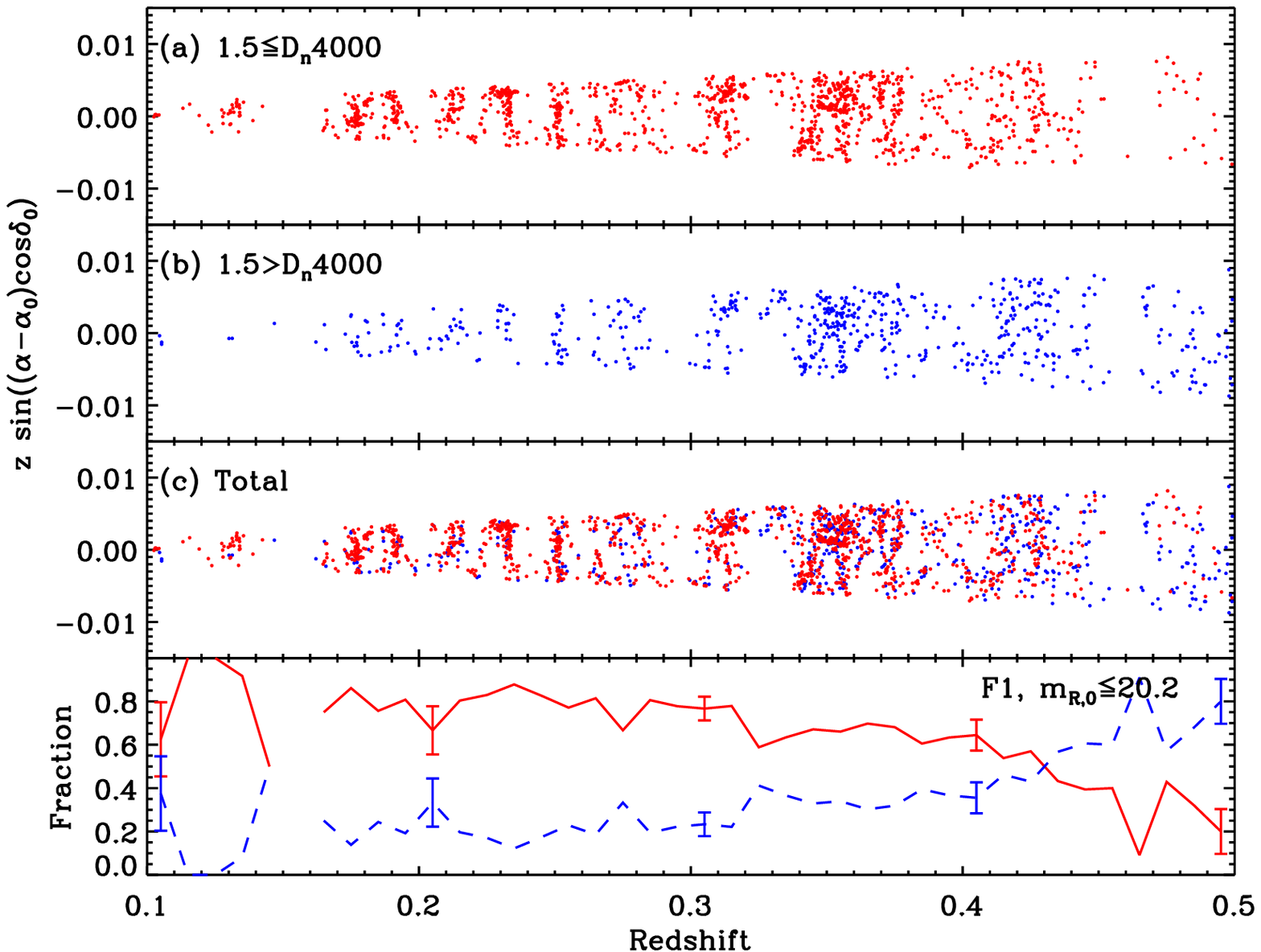}
\includegraphics[width=3.2in]{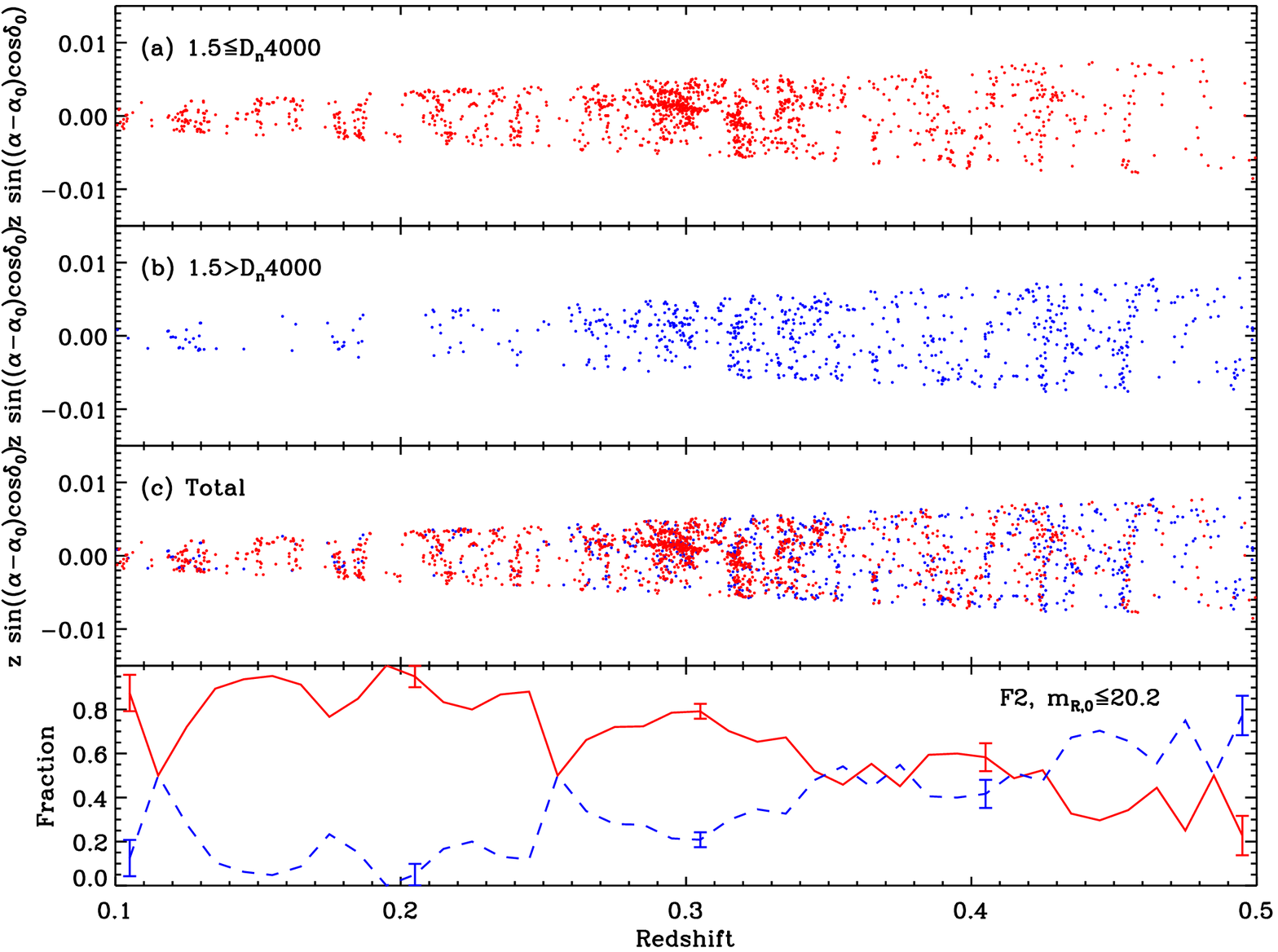}
\end{tabular}
\caption{Cone diagrams projected in R.A. for F1 (left) and F2 (right). We display only galaxies in the stellar mass range 10$^{10}$ - 10$^{11}$ M$_\odot$. Starting from the top the panels for each survey show (1) the cone diagram for galaxies with 1.5$\leq$ D$_n$4000,
(2) the cone diagram for galaxies with 1.5$> $ D$_n$4000, (3) the full cone diagram and (4) the corresponding fractions of quiescent (red) and star-forming (blue) galaxies as a function of redshift.The error bars in the lower panels show the typical 1$\sigma$ error in the fraction.}
\label{massivecone}
\end{figure}

Comparison of the D$_n$4000 distributions for the F1 and F2 fields shows that, averaged over large redshift bins that encompass both dense structures and low density regions, the fractions of  quiescent (large D$_n$4000) and star-forming (small D$_n$4000) galaxies are
surprisingly similar in the two fields. The lower panels of Figure \ref{massivecone} show the behavior of the fractions of star-forming and quiescent galaxies in F1 (left) and F2(right) as a function of redshift for galaxies with stellar masses in the range 10$^{10}$ to 10$^{11}$ M$_\odot$. The upper panels show the corresponding cone diagrams. Comparison of the cone diagrams with the fractions in the lower panels shows the expected dominance of star-forming galaxies in low density regions along with the enhancement of quiescent galaxies in dense regions of the survey. The plots also show clearly that the general trend of the star-forming fraction with redshift is essentially the same for the two fields.
This result is similar to the early conclusion of Bundy et al. (2006) who show that the fractions of quiescent and star-forming galaxies in samples drawn from the DEEP2 data (Davis et al. 2003; Newman et al. 2013) are relatively insensitive to selection effects in their survey.
 
Bundy et al. (2006) conclude, as Figure \ref{subdn4000} emphasizes for F1 and F2, that the quiescent fraction rises with decreasing redshift in every mass bin and the star-forming fraction correspondingly declines. Furthermore, the fraction of star-forming galaxies increases as the stellar mass decreases at fixed redshift. 

The D$_n$4000 distributions for F1 and F2 are also consistent with analyses of the zCOSMOS samples (Scoville et al. 2007; Lilly et al. 2009) by Moresco et al. (2010). Based on 1000 early-type galaxies, they show that in the redshift range 0.45$ < z < 1$, D$_n$4000 decreases with redshift in a given stellar mass range. The SHELS survey offers a larger, complete sample of quiescent galaxies overlapping this redshift range and as a platform for extending a more detailed analysis to lower redshift.   

Although we do not pursue the detailed relationship between galaxy properties and environment here, the SHELS data are very well-suited to such investigations.  The redshift survey is dense and complete and the redshift errors are small ($\lesssim $50 km s$^{-1}$). Thus velocity dispersions even in the thin structures typical of the cosmic web
can be measured robustly. With well-controlled galaxy selection, the combination of the F2 and F1 fields covers a range of environments ranging from a dense complex of massive clusters in F2 to the many obvious low density regions. 

\subsection {Galaxy Counts}
\label{variance}
The F1 and F2 fields are currently unique in the completeness of the redshift surveys to the limiting apparent magnitude. It is thus interesting to compare the galaxy and cluster counts in the two fields as a measure of the potential impact of cosmic variance. It is evident from Sections \ref{mzrelation} and \ref{distdn4000} that the impact is negligible on scaling relations like the MZ relation and on population fractions (as opposed to absolute abundances) as a function of redshift.The potential impact of the variance is obviously largest on quantities dependent on an absolute normalization as a function of redshift.

Table \ref{numbers} lists the count of galaxies in the two fields for various samples. The cleanest samples for comparing F1 with F2 are those limited to R$_0$ = 20.2. There are 7261 and 9489
objects in F1 and F2, respectively to this limit. Most of these objects have a redshift, but among those without a redshift, a small fraction are stars. Using the measured fraction of stars among the  objects with spectroscopy in the two fields (3\% were stars in F2 and 4\% were stars in F1) we can reasonably convert the number of photometric objects to the number of probable galaxies brighter than the limit: 7244(F1) and 9481(F2). In the F1 field we 
selected objects from the SDSS and thus we did not remove regions around stars that were saturated in the DLS photometry as we did for F2. Thus the effective area covered by F1 is larger. There may be some diminution in the counts in F1 as a result of this difference in procedure, but based on the observations, this difference should be $\lesssim 5$\%, essentially the ratio of the effective areas in Table \ref{numbers}. Thus the
ratio of counts r$_g$ in the two fields lies in the range   $ 1.31\pm0.02 < r_g < 1.38 \pm0.02$. For the lower limit we simply take the ratio of the observed counts (we assume that undercounting in F1 roughly compensates for the larger areal coverage); for the upper limit
we normalize the observed counts by the relative areas covered by the two surveys thus assuming that any undercounting of galaxies near bright stars in F1 is negligible.

We can also compare the counts of massive galaxies seen throughout the range $ 0 < z < 0.5$,
a volume of $\sim3\times10^6$ Mpc$^3$, 
in each field.  Based on the data in Table \ref{tab20.2} and Table 3 of Geller et al. (2014), the numbers of galaxies with stellar masses in the range 10.5$ < log(M_{star}/M_{\odot}) <$ 11.5 in the two fields are 2848 (F1) and 3315 (F2).  Correcting for the incompleteness and for the small
difference in the relative volumes of the surveys, the ratio of these populations is 1.19$\pm$0.03. This ratio assumes that the counts of these massive objects are unaffected by bright stars in F1 (in other words we simply correct the relative counts by the small difference in the relative volumes of the two surveys (4.2 (F1)/3.98 (F2); Table \ref{numbers}) and by the small difference in incompleteness (0.94 (F1)/0.97 (F2); Table \ref{numbers})). The count  ratio is at best marginally consistent with the range we obtain from the full magnitude limited count. 

The relative counts of massive galaxies are probably a more robust reflection of any difference between the two fields than the raw counts. Because the observing strategies are not identical, there are a number of subtle systematics that probably have a larger effect on the total count. These include failure to include low surface galaxies in F1 and failure to eliminate regions around bright stars. The magnitude transformation 
(Figure \ref {magcomp}) may also introduce systematics. We have checked the impact of the magnitude conversion by examining the central 3.57 deg$^2$ of the F1 field where only  2.2\% of the objects require a magnitude transformation (as opposed to the 9.8\% for the full field). The average number density of galaxies in this
central field is only $\sim$ 1\% less than in the full field we analyze in detail. Finally F2 contains a set of massive clusters at z$\sim$ 0.53 (note the peak in the redshift histogram of Figure \ref{Bnzf1f2}) that also enhance the total count relative to F1. In contrast, the count of massive galaxies concentrates on high surface brightness objects over the redshift range where both fields are well-sampled.

We next compare the count ratios with the
expected impact of cosmic variance on the relative counts in F1 and F2. 
Driver \& Robotham (2010) use SDSS DR7 as a basis for developing a formula for computing the expected cosmic variance for a survey covering volumes $\lesssim$ 10$^7$ Mpc$^3$ (for H$_0$ = 70 km/s/Mpc) with arbitrary shape and depth (their equation (4)).  They apply their approach to surveys including the VVDS (le Fevre et al. 2005; de la Torre et al. 2007) that covers similar volumes in the redshift range of SHELS.  The VVDS is sparse
to the SHELS limiting apparent magnitude.  Thus, for  comparing the  counts in F1 and F2, the analytic approach of Driver \& Robotham (2014) provides a good guide. Technically their approximation applies to galaxies with absolute luminosities M$^*$$\pm$1 (the numbers of galaxies in this magnitude range are similar to the numbers in the mass range we explore). They also show (their Table 1) that their estimates agree well with independent calculations for galaxies of stellar mass $log(M_{star}/M_{\odot})$ = 10.75 (Moster et al. 2010), analogous to our massive galaxy sample.

The formula given by Driver \& Robotham (2010) provides the expected cosmic variance as a function of the median redshift transverse length and the median radial depth of the survey. The variance they compute for a 4 square degree field covering 0 $ < z < 0.5$ with a 1:1 aspect ratio is  10\%  (their Table 2). From their
equation (4) we derive an identical result for a 4 square degree region with the median 
redshift, 0.29, that we measure for F2 limited to z = 0.5.  For a field like our
4.2 square degree field (F1)  with its slightly lower median redshift, 0.27, for the sample limited to z =0.5, the predicted cosmic variance is also 10\%. 

Driver \& Robotham (2010) show that the distribution of galaxy counts in cells lies between a Gaussian and a log normal distribution. They use the Gaussian approximation in their exploration of variance issues and we follow the same route. If we draw two independent samples randomly from a Gaussian with dispersion $\sigma$, the mean of the absolute difference is (1.13$\pm$ 0.85)$\sigma$. The result is based on 100,000 draws.  The mean
of the absolute difference that we obtain numerically agrees with the analytic prediction,
(2/$\sqrt\pi$)$\sigma$.

For a 10\% cosmic variance ($\sigma$), the observed fractional difference between
between F2 and F1   differs from the simulated mean (1.13$\sigma$ = 0.113) 
lies in the range $0.9 - 3.1$ times the expected error in the mean (0.85$\sigma$ = 0.085). The lower limit corresponds to the count of bright galaxies and the upper limit derives from the raw count ratios. Thus the fractional difference in the abundance of massive galaxies (and the fractional difference in counts to the magnitude limit) is probably consistent with
the expected impact of cosmic variance to within our ability to control for the differing systematics in the two surveys.

\subsection {Massive Weak Lensing Clusters in F1 and F2}
\label{clusters}
The original goal of the SHELS survey of the DLS F1 and F2 fields was comparison of the matter distribution traced by galaxies in a redshift survey with the matter distribution inferred from a weak lensing map. In the F2 field we cross-correlated the galaxy map with the lensing map to show that indeed the lensing map images the projected matter density traced by galaxies in the foreground redshift survey (Geller et al. 2005). Later we examined the correspondence of weak lensing peaks with clusters of galaxies identified from the redshift survey (Geller et al. 2010; Utsumi et al. 2014) and from an x-ray survey (Starikova et al. 2012). These studies underscored some of the difficulties in constructing cluster catalogs. In particular they showed that a threshold signal-to-noise of 4.5 for
detection of a weak lensing peak yields an essentially false positive free catalog. Here we compare weak lensing cluster detections in F2 and F1 for the redshift range of maximum sensitivity, 0.15 $ < z < 0.5$

The F2 field is distinctive because nearly all of the spectroscopic and weak lensing cluster candidates have been observed in the x-ray with either {\it Chandra} or 
{\it XMM-Newton}. Approximately 75\% of the clusters identified from optical spectroscopy are extended x-ray sources; 60\% of the weak lensing peaks with a signal-to-noise of 3.5
correspond to extended x-ray sources. In the redshift range 0.15 $ < z < 0.5$ where clusters are well-sampled in the redshift survey and where the weak lensing map has maximum sensitivity, there are 4 clusters identified cleanly and independently by all three methods: weak lensing, x-ray, and SHELS spectroscopy (these clusters are X0918+2953, X0920+3028, X0920+3030, and X0921+3027 in Table 1 of Starikova et al (2014)). Their inferred masses (M$_{500}$)
range from 7$\times10^{13}$ to 5$\times10^{14}$ M$_\odot$; the corresponding rest frame 
line-of-sight velocity dispersions range from 657$\pm$93 km s$^{-1}$ to
997$\pm$120 km s$^{-1}$.
Figure \ref{maps} shows the positions of the four clusters in F2. They coincide with the
regions on the sky most densely populated by galaxies brighter than our apparent magnitude limit.

In F1, the observations are much less extensive. There are no x-ray observations of extended sources in this field. Ascaso et al. (2014) identified no weak lensing peaks exceeding the detection threshold. However, the DLS data for F1 are of  poorer quality than those for F2; the typical seeing was 0.98$^{\prime\prime}$ as opposed to 0.90$^{\prime\prime}$ for F2. 

\begin{figure}
\centerline{\includegraphics[width=7.0in]{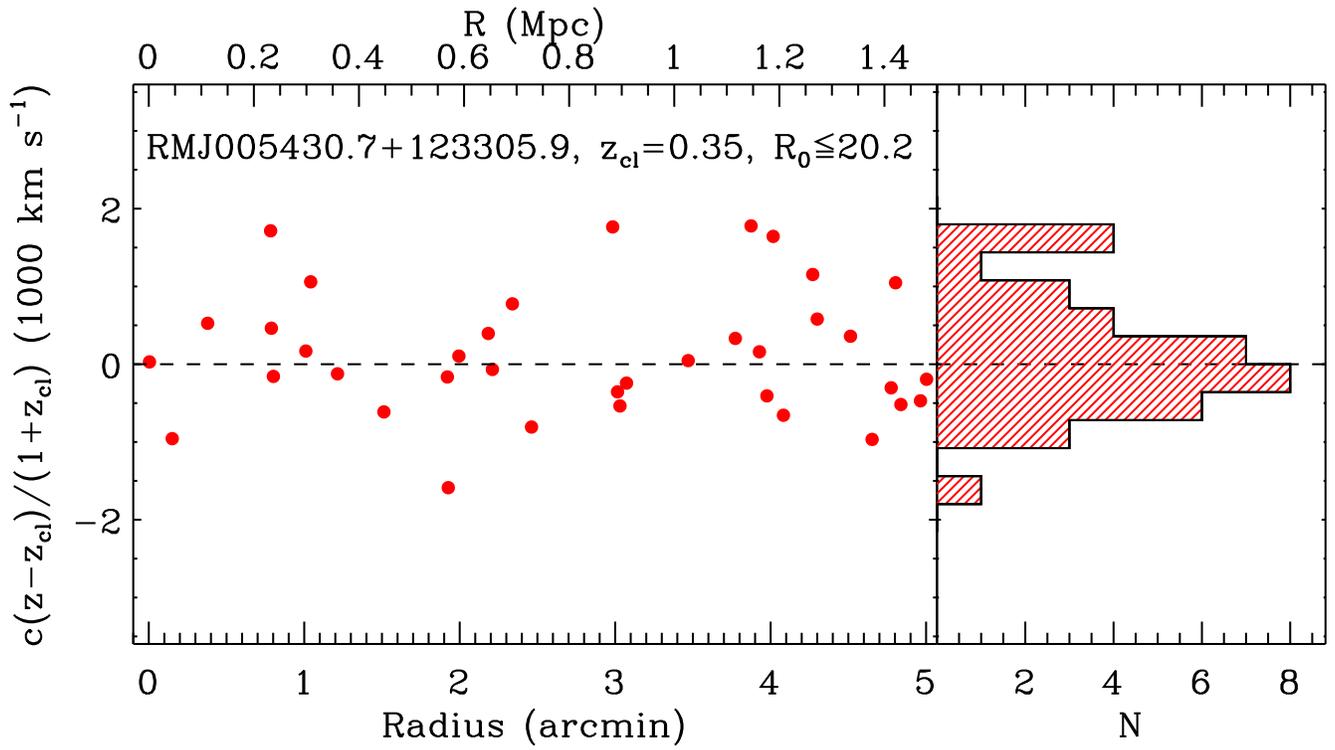}}
\vskip 5ex
\caption{The cluster RMJ005430.7+123305.9 in redshift space. The vertical axis is the rest-frame vleocity relative to the cluster center. The horizontal axis is the projected separation. Note the isolation of the cluster in redshift space. The histogram on the right shows the distribution of rest-frame velocities relative to the cluster center.
\label{rvdiag}}
\end{figure}

In the redshift survey, we identify a single potentially massive cluster in the redshift
range $0.15 < z < 0.5$. This cluster appears as a finger centered at z = 0.35 in the
cone diagram of Figure \ref{BraFcone}. Its position corresponds to the largest peak in the galaxy number density map of Ascaso et al. (2014: Figure A1). In the Ascaso et al. (2014) map, this feature is the only case where a low significance weak lensing peak overlies the peak in the galaxy surface number density. The cluster BCG is located at $\alpha$ = 13.628$^\circ$, $\delta$ = 12.552$^\circ$ and z = 0.35. The SHELS survey contains 37 cluster members with a rest frame line-of-sight velocity dispersion of 834$\pm$126 km s$^{-1}$ (Figure \ref{rvdiag}) within
a projected radius of 1.5 Mpc. The cluster is isolated in redshift space and we make no attempt here to refine the membership. Obviously, tighter limits on the rest-frame velocity would result in a smaller effective line-of-sight velocity dispersion. This cluster is coincident on the sky with
a RedMapper cluster (Rykoff et al. 2014), RMJ005430.7+123305.9; however, the RedMapper photometric redshift is 0.37, inconsistent with our data.

A cluster like RMJ005430.7+123305.9 would probably be detected as a weak lensing peak in F2
at well above the $3\sigma$ level (see Geller et al. 2010; Figure 12).
In F1, however, the poorer image quality implies that a detection would be marginal.
The  worse  and more variable seeing along with the greater Galactic extinction toward F1
degrade the F1 lensing map relative to the analogous map for F2.  Furthermore scattered light from stars outside the field created diffuse sprays of light in the poorly baffled camera during the F1 observations. These effects reduce the effective number of resolved sources by a factor 
of 2.1 per unit area relative to F2. This difference substantially reduces the detection limits for weak lensing.  More precisely,  in F2 clusters with rest frame line-of-sight velocity dispersions in the range 603-697 km s$^{-1}$ would be weak lensing detections at a signal-to-noise of 3-4 (Geller et al. 2010); in F1 the corresponding range computed as in Geller et al. (2010) is
753-870 km s$^{-1}$.  This reduction in sensitivity probably accounts for the absence of a weak lensing  detection by Ascaso et al. (2014).

We conclude that although the difference between F1 and F2 seems remarkable at first glance, the relative galaxy counts are  consistent with  cosmic variance estimates. The difference in the number of clusters detected  in the weak lensing maps is driven primarily by shot noise along with significant differences in the observing conditions that affect the construction of a weak lensing map for F1.

\section{Conclusion}

The SHELS project covers two widely separated 4 square degree fields of the Deep Lens Survey.
These fields include 16,055 redshifts and they are both $\gtrsim$ 94\% complete
to an extinction corrected R$_0$ = 20.2. Other surveys covering comparable areas to   the same apparent magnitude limit are sparser and/or color-selected. The straightforward selection in apparent magnitude makes SHELS a useful benchmark for evaluating selection effects based on other approaches.

The median redshift of both SHELS fields is z $\sim$ 0.3. The redshift histograms of the two regions nonetheless differ significantly. This difference is largely driven by differences in the details of the large-scale structure in each region, particularly the presence of 
several massive clusters of galaxies in F2.

We have previously used the SHELS data to determine the stellar mass-metallicity relation 
for the redshift interval 0.2$ < z < 0.38$. Comparison of the relation determined for the F1 and F2 fields separately shows that the fiducial mass, M$_0$ characterizing the relation changes insignificantly (at the 1.6 $\sigma$ level) from one field to the other. The relation for the two fields is a remarkably robust estimate for this redshift range. We provide the metallicities and stellar masses as we
did for F2 in Geller et al. (2014).

As we did for the F2 field (Geller et al. 2014) we use the distribution of the
spectral indicator D$_n$4000 as a proxy to discriminate between the star-forming and quiescent populations as a function of redshift and stellar mass. The behavior of the D$_n$4000 distributions for the two fields is remarkably similar with salient differences driven
either by low redshift structures sampled to low stellar mass or to the presence of massive clusters of galaxies. For galaxy stellar masses in the range 10$^{10}$ to 10$^{11}$M$_\odot$, the star-forming population fraction as a function of redshift are remarkably similar for the two fields in spite of the difference in the overall mean galaxy density. These
broadly binned results  are 
a guide to the use of the data for more detailed analysis of the properties of galaxies as a function of their environment. As a result of the small redshift errors ($\lesssim 50 $km s$^{-1}$), the SHELS data are particularly well-suited to this task. 

In contrast with the stable population fractions, the raw counts of galaxies and massive clusters in the two fields seem, at first glance, to differ significantly.  The procedures in observing the two fields were not identical thus complicating the comparison.   However, the
31-38\% difference in the counts to the limiting magnitude (and the smaller 19\% in the  count of massive galaxies) is probably consistent with the expected cosmic variance to within our ability to control for the relative systematics in the two surveys.

Comparing the abundance of massive clusters in the two fields is complicated by the much poorer data available for F1. In the F2 field there are extensive x-ray observations and the weak lensing data is of higher quality than for F1. In fact, the SHELS survey uncovers a cluster in F1 at z = 0.35 that probably should have been detected in F1 if the lensing data were of the same quality as for the F2 field. The difference in the cluster count is dominated by shot noise, but the apparent difference is accentuated by the poorer F1 lensing data. This comparison underscores the need for well-controlled calibration of surveys and underscores
the subtle issues that enter the comparison of data for fields observed under different conditions
and with different observational approaches.

The SHELS survey covers 8 square degrees in two widely separated fields and includes 20,754 redshifts for galaxies with R$\leq$ 20.6 along with 4457 redshifts for fainter objects.
The complete surveys of the F1 and F2 fields provide a resource for many investigations of galaxy properties and their environments. The completeness of the surveys to the apparent magnitude limit provides benchmarks for color-selected surveys and for the development of new strategies based on combinations of imaging and spectroscopy.

\begin{acknowledgments}
We appreciate a thoughtful, careful review by the referee that substantially improved this paper. We thank Changbom Park of KIAS for generous support that enabled completion of this project. We thank Kairy Herrera of Brown University for examining images of faint galaxies with superimposed stars to refine the construction of Table 4. We thank Warren Brown for remeasuring two redshifts for two of the objects examined by Kairy Herrera.
Scott Kenyon provided advice on numerical methods along with comments on the manuscript.  
Perry Berlind and Mike Calkins masterfully operated the Hectospec. Susan Tokarz and Sean Moran assisted with the data reduction. The Smithsonian Institution supported the
research of Margaret Geller, Ho Seong Hwang, Michael Kurtz, Daniel Fabricant and Jabran Zahid. KIAS supported the research of Ho Seong Hwang.
\end{acknowledgments}

{\it Facilities:}\facility {MMT(Hectospec)}

%%%%%%%%%%%%%%%%%%%%%%%%%%%%%%%%%%%%%%%%%%%%%%%%%%%%%%%%%%%%%%%%%%%%%%%%%%%%%%%
%%%%%%%%%%%%%%%%%%%%%%%%         table 1      %%%%%%%%%%%%%%%%%%%%%%%%%%%%%%%%
%%%%%%%%%%%%%%%%%%%%%%%%%%%%%%%%%%%%%%%%%%%%%%%%%%%%%%%%%%%%%%%%%%%%%%%%%%%%%%%
\begin{deluxetable}{ccccccccrc}
\rotate
\tabletypesize{\footnotesize}
\tablewidth{0pc} %\tablenum{2}
\tablecaption{SHELS Redshifts with $m_{R,0}\leq20.2$
\label{tab20.2}}
\tablehead{
SHELS ID & SDSS ObjID & $m_{R,0}$& $z$ & $z$                     &$m_{R,0}$               & 
Flag\tablenotemark{d}& $D_n4000$ & log($M_\star/M_\odot)$ & 12+        \\
         &            & (mag)    &     & Source\tablenotemark{b} &Source\tablenotemark{c} &  
                    &           &                        & log(O/H)
}
\startdata
   12.324533+12.480500 & 1237678919673053653 & $19.881\pm0.095$ & $ 0.40251\pm0.00015$ & 1 & 
2 & 0 & 1.76 &   $10.85^{+0.11}_{-0.19}$ &   ... \\
   12.324566+12.653818 & 1237678859550589594 & $18.646\pm0.058$ & $ 0.25096\pm0.00006$ & 1 & 
2 & 0 & 1.23 &   $10.33^{+0.19}_{-0.34}$ &  9.05 \\
   12.324614+12.798466 & 1237678859550589182 & $19.893\pm0.192$ & $ 1.46612\pm0.00033$ & 1 & 
2 & 0 &  ... &   $ 8.86^{+0.17}_{-0.19}$ &   ... \\
   12.325052+11.897449 & 1237678858476913241 & $20.125\pm0.165$ & $ 0.55024\pm0.00022$ & 1 & 
2 & 0 & 1.82 &   $11.20^{+0.11}_{-0.13}$ &   ... \\
   12.325632+12.798472 & 1237678859550589183 & $18.679\pm0.119$ & $ 0.19739\pm0.00017$ & 1 & 
2 & 0 & 1.28 &   $10.01^{+0.14}_{-0.10}$ &   ... \\
   12.327368+12.549498 & 1237678919673053413 & $19.346\pm0.288$ & $ 0.27016\pm0.00014$ & 1 & 
2 & 0 & 1.96 &   $10.66^{+0.11}_{-0.17}$ &   ... \\
   12.327427+12.550200 & 1237678919673053412 & $18.050\pm0.528$ & $ 0.26629\pm0.00011$ & 1 & 
2 & 0 & 1.78 &   $11.36^{+0.07}_{-0.08}$ &   ... \\
   12.327430+11.594275 & 1237678918599311681 & $19.955\pm0.078$ & $ 0.27939\pm0.00007$ & 1 & 
2 & 0 & 1.01 &   $ 9.48^{+0.14}_{-0.14}$ &  8.77 \\
   12.327575+12.491597 & 1237678919673053661 & $19.978\pm0.081$ & $ 0.26447\pm0.00010$ & 1 & 
2 & 0 & 1.40 &   $ 9.82^{+0.19}_{-0.27}$ &  9.14 \\
   12.327611+12.244888 & 1237678859013783788 & $19.887\pm0.114$ & $ 0.49797\pm0.00009$ & 1 & 
2 & 0 & 1.22 &   $10.56^{+0.16}_{-0.13}$ &   ... \\
\enddata
\tablenotetext{a}{This table is available in its entirety in a machine-readable form in the 
online journal. A portion is shown here for guidance regarding its form and content.}
\tablenotetext{b}{(1) This study; (2) SDSS; (3) NED.}
\tablenotetext{c}{(1) DLS; (2) SDSS.}
\tablenotetext{d}{(0) Extended source; (1) Point source.}
\end{deluxetable}

\clearpage

%%%%%%%%%%%%%%%%%%%%%%%%%%%%%%%%%%%%%%%%%%%%%%%%%%%%%%%%%%%%%%%%%%%%%%%%%%%%%%%
%%%%%%%%%%%%%%%%%%%%%%%%         table 2      %%%%%%%%%%%%%%%%%%%%%%%%%%%%%%%%
%%%%%%%%%%%%%%%%%%%%%%%%%%%%%%%%%%%%%%%%%%%%%%%%%%%%%%%%%%%%%%%%%%%%%%%%%%%%%%%
\begin{deluxetable}{ccccccccrc}
\rotate
\tabletypesize{\footnotesize}
\tablewidth{0pc} %\tablenum{2}
\tablecaption{SHELS Redshifts with $m_{R,0}>20.2$
\label{tabfaint}}
\tablehead{
SHELS ID & SDSS ObjID & $m_{R,0}$& $z$ & $z$                     &$m_{R,0}$               & 
Flag\tablenotemark{d}& $D_n4000$ & log($M_\star/M_\odot)$ & 12+        \\
         &            & (mag)    &     & Source\tablenotemark{b} &Source\tablenotemark{c} &  
                    &           &                        & log(O/H)
}
\startdata
   12.389126+13.290399 & 1237678920746795438 & $20.495\pm0.007$ & $ 0.27409\pm0.00002$ & 1 & 
1 & 0 & 1.34 &   $ 9.32^{+0.22}_{-0.21}$ &   ... \\
   12.389223+13.548410 & 1237678860624396698 & $20.591\pm0.132$ & $ 0.34567\pm0.00011$ & 1 & 
2 & 0 & 1.71 &   $10.31^{+0.15}_{-0.20}$ &   ... \\
   12.389453+13.234157 & 1237678860087525736 & $20.474\pm0.006$ & $ 0.33834\pm0.00013$ & 1 & 
1 & 0 & 1.92 &   $10.58^{+0.17}_{-0.18}$ &   ... \\
   12.390178+12.722615 & 1237678859550654860 & $20.270\pm0.007$ & $ 0.43582\pm0.00019$ & 1 & 
1 & 0 & 1.16 &   $10.31^{+0.31}_{-0.20}$ &   ... \\
   12.391906+12.125328 & 1237678919136182371 & $20.622\pm0.005$ & $ 0.63710\pm0.00019$ & 1 & 
1 & 0 & 1.36 &   $10.81^{+0.25}_{-0.32}$ &   ... \\
   12.392926+13.139955 & 1237678860087525754 & $20.627\pm0.007$ & $ 0.52116\pm0.00011$ & 1 & 
1 & 0 & 1.43 &   $10.72^{+0.21}_{-0.17}$ &   ... \\
   12.396500+12.704612 & 1237678859550654877 & $20.475\pm0.007$ & $ 0.34583\pm0.00007$ & 1 & 
1 & 0 & 1.15 &   $ 9.50^{+0.20}_{-0.18}$ &  8.77 \\
   12.396858+13.245761 & 1237678920746795610 & $20.300\pm0.007$ & $ 0.52204\pm0.00019$ & 1 & 
1 & 0 & 1.34 &   $10.74^{+0.22}_{-0.25}$ &   ... \\
   12.397086+12.221524 & 1237678859013783938 & $20.205\pm0.006$ & $ 0.55371\pm0.00019$ & 1 & 
1 & 0 & 1.79 &   $10.89^{+0.25}_{-0.25}$ &   ... \\
   12.397125+12.852595 & 1237678920209924690 & $20.555\pm0.011$ & $ 0.48872\pm0.00011$ & 1 & 
1 & 0 & 1.31 &   $10.09^{+0.23}_{-0.21}$ &   ... \\
\enddata
\tablenotetext{a}{This table is available in its entirety in a machine-readable form in the 
online journal. A portion is shown here for guidance regarding its form and content.}
\tablenotetext{b}{(1) This study; (2) SDSS; (3) NED.}
\tablenotetext{c}{(1) DLS; (2) SDSS.}
\tablenotetext{d}{(0) Extended source; (1) Point source.}
\end{deluxetable}

\clearpage
%%%%%%%%%%%%%%%%%%%%%%%%%%%%%%%%%%%%%%%%%%%%%%%%%%%%%%%%%%%%%%%%%%%%%%%%%%%%%%%
%%%%%%%%%%%%%%%%%%%%%%%%         table 3       %%%%%%%%%%%%%%%%%%%%%%%%%%%%%%%%
%%%%%%%%%%%%%%%%%%%%%%%%%%%%%%%%%%%%%%%%%%%%%%%%%%%%%%%%%%%%%%%%%%%%%%%%%%%%%%%
\begin{deluxetable}{ccc}
\tabletypesize{\footnotesize}
\tablewidth{0pc} %\tablenum{2}
\tablecaption{Objects without redshifts at $m_{R,0}\leq20.2$\tablenotemark{a,b}
\label{tabnoz}}
\tablehead{
SHELS ID & SDSS ObjID & $m_{R,0}$   \\
         &            & (mag) 
}
\startdata
   12.315080+12.318782 & 1237678859013718622 & $19.878\pm0.133$ \\
   12.339393+12.349747 & 1237678859013783808 & $20.108\pm0.330$ \\
   12.352705+13.175315 & 1237678860087525659 & $20.175\pm0.266$ \\
   12.358372+11.543474 & 1237678857940042136 & $20.140\pm0.110$ \\
   12.359576+12.377343 & 1237678859013783569 & $20.098\pm0.009$ \\
   12.360730+12.853304 & 1237678920209924618 & $19.715\pm0.078$ \\
   12.363524+13.473994 & 1237678860624396650 & $20.131\pm0.146$ \\
   12.363848+12.351914 &                 ... & $19.445\pm0.007$ \\
   12.365371+12.855583 & 1237678920209924173 & $19.781\pm0.139$ \\
   12.367078+12.860638 & 1237678920209924839 & $20.152\pm0.235$ \\
\enddata
\tablenotetext{a}{This table is available in its entirety in a machine-readable form in the 
online journal. A portion is shown here for guidance regarding its form and content.}
\tablenotetext{b}{The symbol ... in the SDSS ObjID column signified that the object is in the DLS
catalog but absent from the SDSS. There are 20 of these objects in the catalog.}
\end{deluxetable}

\clearpage
%%%%%%%%%%%%%%%%%%%%%%%%%%%%%%%%%%%%%%%%%%%%%%%%%%%%%%%%%%%%%%%%%%%%%%%%%%%%%%%
%%%%%%%%%%%%%%%%%%%%%%%%         table 4       %%%%%%%%%%%%%%%%%%%%%%%%%%%%%%%%
%%%%%%%%%%%%%%%%%%%%%%%%%%%%%%%%%%%%%%%%%%%%%%%%%%%%%%%%%%%%%%%%%%%%%%%%%%%%%%%
\begin{deluxetable}{cccccc}
\rotate
\tabletypesize{\footnotesize}
\tablewidth{0pc} %\tablenum{2}
\tablecaption{SHELS Redshifts with $z<0.0015$
\label{tabstar}}
\tablehead{
SHELS ID & SDSS ObjID & $m_{R,0}$& $z$ & $z$                     &$m_{R,0}$  \\
         &            & (mag)    &     & Source\tablenotemark{b} &Source\tablenotemark{c}
}
\startdata
   12.311022+13.024869 & 1237678920209924333 & $18.017\pm0.011$ & $-0.00069\pm0.00001$ & 2 & 
2 \\
   12.322512+12.238044 & 1237678859013783624 & $17.262\pm0.008$ & $-0.00005\pm0.00006$ & 1 & 
2 \\
   12.326695+12.168094 & 1237678919136182450 & $18.425\pm0.015$ & $-0.00026\pm0.00005$ & 1 & 
2 \\
   12.331113+13.070127 & 1237678860087460426 & $20.031\pm0.065$ & $-0.00009\pm0.00021$ & 1 & 
2 \\
   12.331615+12.833143 & 1237678920209924451 & $19.304\pm0.158$ & $-0.00074\pm0.00006$ & 2 & 
2 \\
   12.334228+12.140061 & 1237678919136182455 & $17.407\pm0.011$ & $ 0.00011\pm0.00001$ & 2 & 
2 \\
   12.339199+11.762720 & 1237678918599311568 & $20.611\pm0.008$ & $-0.00034\pm0.00012$ & 1 & 
1 \\
   12.343303+12.813752 & 1237678859550654681 & $20.244\pm0.004$ & $-0.00002\pm0.00012$ & 1 & 
1 \\
   12.350332+11.883782 & 1237678858476912807 & $19.114\pm0.002$ & $-0.00008\pm0.00021$ & 1 & 
1 \\
   12.355689+12.097490 & 1237678919136182468 & $18.571\pm0.001$ & $-0.00051\pm0.00007$ & 1 & 
1 \\
\enddata
\tablenotetext{a}{This table is available in its entirety in a machine-readable form in the 
online journal. A portion is shown here for guidance regarding its form and content.}
\tablenotetext{b}{(1) This study; (2) SDSS}
\tablenotetext{c}{(1) DLS; (2) SDSS.}
\end{deluxetable}

\clearpage
%%%%%%%%%%%%%%%%%%%%%%%%%%%%%%%%%%%%%%%%%%%%%%%%%%%%%%%%%%%%%%%%%%%%%%%%%%%%%%%
%%%%%%%%%%%%%%%%%%%%%%%%         table 5      %%%%%%%%%%%%%%%%%%%%%%%%%%%%%%%%
%%%%%%%%%%%%%%%%%%%%%%%%%%%%%%%%%%%%%%%%%%%%%%%%%%%%%%%%%%%%%%%%%%%%%%%%%%%%%%%
\begin{deluxetable}{lccr}
\tabletypesize{\normalsize}
\tablewidth{0pc}
\tablecaption{\large SHELS F1 and F2 Redshift Survey Properties
\label{numbers}}
\tablehead{
Parameter             & Value (F1)        & Value (F2)}
\startdata
E(B$-$V)              & 0.06 (A$_R$=0.16) & 0.02 (A$_R$=0.05) \\
Survey Area (deg$^2$) & 4.20& 3.98 (excludes masked area)\\
\hline
N$_{phot, R_0\leq20.2}$\tablenotemark{a}  &  7261 & 9489\\
N$_{z, R_0\leq20.2}$\tablenotemark{b}     &  6839 &  9216\\
Completeness$_{20.2}$                     & 94.2\% & 97.1\%\\
z$_{med, R_0\leq20.2}$                    & 0.282 &  0.294\\
\hline
N$_{phot, 20.3}\tablenotemark{c}$         &  6626 &   9946\\
N$_{z, 20.3}\tablenotemark{d}$            &  6345 &   9643\\ 
Completeness$_{20.3}$                     & 95.8\%  & 97.0\%\\
\hline
N$_{phot, 20.6}\tablenotemark{c}$         &  9207 &  13408\\
N$_{z, 20.6}\tablenotemark{d}$            &  8049 &  12705\\
Completeness$_{20.6}$                     & 87.4\%  & 94.8\%\\
\hline
N$_{z, R > 20.6}$\tablenotemark{c}        &  1515 &   2942\\
N$_{z, total}$           &  9564\tablenotemark{e} &  16319\tablenotemark{f}\\
\enddata
\tablenotetext{a}{Number of photometric objects in the complete survey region brighter than 
the quoted limit R$_0$ corrected for Galactic extinction.}
\tablenotetext{b}{Number of redshifts in the complete survey region brighter than the quoted 
limit R$_0$ corrected for Galactic extinction.}
\tablenotetext{c}{Number of photometric objects brighter than the specified uncorrected DLS 
R-band limit in the complete survey region.}
\tablenotetext{d}{Number of redshifts brighter than the specified uncorrected DLS R-band 
limit in the complete survey region.}
\tablenotetext{e}{In addition the spectroscopy in F1 identified 413 stars (misclassified by 
SDSS as galaxies) and 297 QSOs (point sources in SDSS).}
\tablenotetext{f}{All F2 redshifts published in Geller at al. (2014).}
\end{deluxetable}

\clearpage
%%%%%%%%%%%%%%%%%%%%%%%%%%%%%%%%%%%%%%%%%%%%%%%%%%%%%%%%%%%%%%%%%%%%%%%%%%%%%%%
%%%%%%%%%%%%%%%%%%%%%%%%         table 6      %%%%%%%%%%%%%%%%%%%%%%%%%%%%%%%%
%%%%%%%%%%%%%%%%%%%%%%%%%%%%%%%%%%%%%%%%%%%%%%%%%%%%%%%%%%%%%%%%%%%%%%%%%%%%%%%
\begin{deluxetable}{lccl}
\tabletypesize{\normalsize}
\tablewidth{0pc} %\tablenum{2}
\setlength{\tabcolsep}{0.2in}
\tablecaption{\large Mass-Metallicity Relations for F1 and F2\tablenotemark{a}
\label{tblMZ}}
\tablehead{
Parameter & Value (F1)& Value (F2) &Value (F1+F2)\tablenotemark{b}}
\startdata
   N$_{MZ}$\tablenotemark{c}&1446&2131&3577\\
   log(M$_0$/M$_\odot$)& 9.56$\pm$0.03&9.50$\pm$0.02&9.52$\pm$0.02\\
   Z$_0$ & 9.10$\pm$0.01& 9.10$\pm$0.01&9.10$\pm$0.004\\
   
   $\gamma$&0.49$\pm$0.03&0.52$\pm$0.03&0.52$\pm$0.02\\
   
\enddata
\tablenotetext{a}{Errors are 1$\sigma$ bootstrap errors throughout the Table.}
\tablenotetext{b} {Values are from line 2 (best fit) of Table 2, Zahid et al. (2014)}
\tablenotetext{c} {Number of star forming objects included in the MZ relation. The total
is identical to the nubmer of objects in Zahid et al.(2014)}. 
\end{deluxetable}
\clearpage

\clearpage
\clearpage

\clearpage

\clearpage

\clearpage

%\begin{figure}
%\centerline{\includegraphics[width=7.0in]{f6.eps}}
%\vskip 5ex
%\caption{
%} 
%\label{Fcone}
%\end{figure}\clearpage

\clearpage

\clearpage

\clearpage

%\begin{figure}
%\centerline{\includegraphics[width=7.0in]{f8.eps}}
%\vskip 5ex
%\caption{Observed $(g-r)_{fiber,0}$ and $(r-i)_{fiber,0}$ SDSS extinction-corrected colors for the F1 sample (left panels). (Right) K-corrected colors for F1 galaxies shifted to
%$ z = 0.35$. CORRECT?? We display only CORRECT?? 30\% of the data for clarity. The narrowing color distribution at the largest redshifts
%occurs because the magnitude limited sample contains increasingly luminous galaxies that are generally somewhat redder.
%\label{BFzcol}}
%\end{sidewaysfigure}
%\end{figure}
%\clearpage
\clearpage

\clearpage

\clearpage

\clearpage


\clearpage
\begin{thebibliography}


\bibitem[Abell(1958)]{1958ApJS....3..211A} Abell, G.~O.\ 1958, \apjs, 3, 
211 

\bibitem[Ahn et al.(2014)]{2014ApJS..211...17A} Ahn, C.~P., Alexandroff, 
R., Allende Prieto, C., et al.\ 2014, \apjs, 211, 17 

\bibitem[Alam et al.(2015)]{2015ApJS..219...12A} Alam, S., Albareti, F.~D., 
Allende Prieto, C., et al.\ 2015, \apjs, 219, 12 


\bibitem[Arnouts et al.(1999)]{1999MNRAS.310..540A} Arnouts, S., Cristiani, 
S., Moscardini, L., et al.\ 1999, \mnras, 310, 540 


\bibitem[Ascaso et al.(2014)]{2014MNRAS.439.1980A} Ascaso, B., Wittman, D., 
\& Dawson, W.\ 2014, \mnras, 439, 1980 

\bibitem[Baldwin et al.(1981)]{1981PASP...93....5B} Baldwin, J.~A., 
Phillips, M.~M., \& Terlevich, R.\ 1981, \pasp, 93, 5 


\bibitem[Balogh et al.(1999)]{1999ApJ...527...54B} Balogh, M.~L., Morris, 
S.~L., Yee, H.~K.~C., Carlberg, R.~G., 
\& Ellingson, E.\ 1999, \apj, 527, 54 


\bibitem[Bell et al.(2003)]{2003ApJS..149..289B} Bell, E.~F., McIntosh, 
D.~H., Katz, N., \& Weinberg, M.~D.\ 2003, \apjs, 149, 289 

\bibitem[Bielby et 
al.(2014)]{2014A&A...568A..24B} Bielby, R.~M., Gonzalez-Perez, V., McCracken, H.~J., et al.\ 2014, \aap, 568, A24 

\bibitem[Blanton et al.(2005)]{2005ApJ...631..208B} Blanton, M.~R., Lupton, 
R.~H., Schlegel, D.~J., et al.\ 2005, \apj, 631, 208 


\bibitem[Brinchmann 
\& Ellis(2000)]{2000ApJ...536L..77B} Brinchmann, J., \& Ellis, R.~S.\ 2000, \apjl, 536, L77 






\bibitem[Bundy et al.(2006)]{2006ApJ...651..120B} Bundy, K., Ellis, R.~S., 
Conselice, C.~J., et al.\ 2006, \apj, 651, 120 

\bibitem[Chang et al.(2015)]{2015PhRvL.115e1301C} Chang, C., Vikram, V., 
Jain, B., et al.\ 2015, Physical Review Letters, 115, 051301 

\bibitem[Damjanov et al.(2015)]{2015arXiv150803346D} Damjanov, I., Zahid, 
H.~J., Geller, M.~J., \& Hwang, H.~S.\ 2015, arXiv:1508.03346 



\bibitem[Davis et al.(2003)]{2003SPIE.4834..161D} Davis, M., Faber, S.~M., 
Newman, J., et al.\ 2003, \procspie, 4834, 161 

\bibitem[de la Torre et 
al.(2007)]{2007A&A...475..443D} de la Torre, S., Le F{\`e}vre, O., Arnouts, S., et al.\ 2007, \aap, 475, 443 

\bibitem[Driver 
\& Robotham(2010)]{2010MNRAS.407.2131D} Driver, S.~P., \& Robotham, A.~S.~G.\ 2010, \mnras, 407, 2131 

\bibitem[Fabricant et al.(1998)]{1998SPIE.3355..285F} Fabricant, D.~G., 
Hertz, E.~N., Szentgyorgyi, A.~H., et al.\ 1998, \procspie, 3355, 285 


\bibitem[Fabricant et al.(2005)]{2005PASP..117.1411F} Fabricant, D., Fata, 
R., Roll, J., et al.\ 2005, \pasp, 117, 1411 

\bibitem[Fabricant et al.(2008)]{2008PASP..120.1222F} Fabricant, D.~G., 
Kurtz, M.~J., Geller, M.~J., et al.\ 2008, \pasp, 120, 1222 



\bibitem[Fabricant et al.(2013)]{2013PASP..125.1362F} Fabricant, D., 
Chilingarian, I., Hwang, H.~S., et al.\ 2013, \pasp, 125, 1362 



\bibitem[Freedman Woods et al.(2010)]{2010AJ....139.1857F} Freedman Woods, 
D., Geller, M.~J., Kurtz, M.~J., et al.\ 2010, \aj, 139, 1857 


\bibitem[Garilli et 
al.(2008)]{2008A&A...486..683G} Garilli, B., Le F{\`e}vre, O., Guzzo, L., et al.\ 2008, \aap, 486, 683 


\bibitem[Geller 
\& Huchra(1989)]{1989Sci...246..897G} Geller, M.~J., \& Huchra, J.~P.\ 1989, Science, 246, 897 




\bibitem[Geller et al.(2005)]{2005ApJ...635L.125G} Geller, M.~J., 
Dell'Antonio, I.~P., Kurtz, M.~J., et al.\ 2005, \apjl, 635, L125 



\bibitem[Geller et al.(2010)]{2010ApJ...709..832G} Geller, M.~J., Kurtz, 
M.~J., Dell'Antonio, I.~P., Ramella, M., 
\& Fabricant, D.~G.\ 2010, \apj, 709, 832 

\bibitem[Geller et al.(2012)]{2012AJ....143..102G} Geller, M.~J., Diaferio, 
A., Kurtz, M.~J., Dell'Antonio, I.~P., 
\& Fabricant, D.~G.\ 2012, \aj, 143, 102 

\bibitem[Geller et al.(2014)]{2014ApJS..213...35G} Geller, M.~J., Hwang, 
H.~S., Fabricant, D.~G., et al.\ 2014, \apjs, 213, 35 

\bibitem[Geller 
\& Hwang(2015)]{2015AN....336..428G} Geller, M.~J., \& Hwang, H.~S.\ 2015, Astronomische Nachrichten, 336, 428 


\bibitem[Hwang et al.(2012)]{2012ApJ...758...25H} Hwang, H.~S., Geller, 
M.~J., Kurtz, M.~J., Dell'Antonio, I.~P., 
\& Fabricant, D.~G.\ 2012, \apj, 758, 25 


\bibitem[Ilbert et 
al.(2006)]{2006A&A...457..841I} Ilbert, O., Arnouts, S., McCracken, H.~J., et al.\ 2006, \aap, 457, 841 



\bibitem[Kauffmann et al.(2003)]{2003MNRAS.341...33K} Kauffmann, G., 
Heckman, T.~M., White, S.~D.~M., et al.\ 2003, \mnras, 341, 33 

\bibitem[Kewley et al.(2005)]{2005PASP..117..227K} Kewley, L.~J., Jansen, 
R.~A., \& Geller, M.~J.\ 2005, \pasp, 117, 227 

\bibitem[Kewley et al.(2006)]{2006MNRAS.372..961K} Kewley, L.~J., Groves, 
B., Kauffmann, G., \& Heckman, T.\ 2006, \mnras, 372, 961 



\bibitem[Kobulnicky 
\& Kewley(2004); KK04]{2004ApJ...617..240K} Kobulnicky, H.~A., \& Kewley, L.~J.\ 2004, \apj, 617, 240 


\bibitem[Kochanek et al.(2012)]{2012ApJS..200....8K} Kochanek, C.~S., 
Eisenstein, D.~J., Cool, R.~J., et al.\ 2012, \apjs, 200, 8 

\bibitem[Kurtz 
\& Mink(1998)]{1998PASP..110..934K} Kurtz, M.~J., \& Mink, D.~J.\ 1998, \pasp, 110, 934 



\bibitem[Le F{\`e}vre et 
al.(2013)]{2013A&A...559A..14L} Le F{\`e}vre, O., Cassata, P., Cucciati, O., et al.\ 2013, \aap, 559, A14 


\bibitem[Lilly et al.(2009)]{2009ApJS..184..218L} Lilly, S.~J., Le Brun, 
V., Maier, C., et al.\ 2009, \apjs, 184, 218 

\bibitem[Mignoli et 
al.(2005)]{2005A&A...437..883M} Mignoli, M., Cimatti, A., Zamorani, G., et al.\ 2005, \aap, 437, 883 

\bibitem[Mink et al.(2007)]{2007ASPC..376..249M} Mink, D.~J., Wyatt, W.~F., 
Caldwell, N., et al.\ 2007, Astronomical Data Analysis Software and Systems 
XVI, 376, 249 



\bibitem[Moresco et 
al.(2010)]{2010A&A...524A..67M} Moresco, M., Pozzetti, L., Cimatti, A., et al.\ 2010, \aap, 524, A67 


\bibitem[Moresco et 
al.(2013)]{2013A&A...558A..61M} Moresco, M., Pozzetti, L., Cimatti, A., et al.\ 2013, \aap, 558, A61 
\bibitem[Moster et al.(2010)]{2010ApJ...710..903M} Moster, B.~P., 
Somerville, R.~S., Maulbetsch, C., et al.\ 2010, \apj, 710, 903 


\bibitem[Moustakas et al.(2013)]{2013ApJ...767...50M} Moustakas, J., Coil, 
A.~L., Aird, J., et al.\ 2013, \apj, 767, 50 




\bibitem[Newman et al.(2013)]{2013ApJS..208....5N} Newman, J.~A., Cooper, 
M.~C., Davis, M., et al.\ 2013, \apjs, 208, 5 



\bibitem[Noeske et al.(2007)]{2007ApJ...660L..47N} Noeske, K.~G., Faber, 
S.~M., Weiner, B.~J., et al.\ 2007, \apjl, 660, L47 

\bibitem[Peng 
\& Maiolino(2014)]{2014MNRAS.438..262P} Peng, Y.-j., \& Maiolino, R.\ 2014, \mnras, 438, 262 


\bibitem[Roll et al.(1998)]{1998SPIE.3355..324R} Roll, J.~B., Fabricant, 
D.~G., \& McLeod, B.~A.\ 1998, \procspie, 3355, 324 



\bibitem[Roseboom et al.(2006)]{2006MNRAS.373..349R} Roseboom, I.~G., 
Pimbblet, K.~A., Drinkwater, M.~J., et al.\ 2006, \mnras, 373, 349 

\bibitem[Rykoff et al.(2014)]{2014ApJ...785..104R} Rykoff, E.~S., Rozo, E., 
Busha, M.~T., et al.\ 2014, \apj, 785, 104 


\bibitem[Scoville et al.(2007)]{2007ApJS..172..150S} Scoville, N., Aussel, 
H., Benson, A., et al.\ 2007, \apjs, 172, 150 

\bibitem[Shan et al.(2012)]{2012ApJ...748...56S} Shan, H., Kneib, J.-P., 
Tao, C., et al.\ 2012, \apj, 748, 56 


\bibitem[Starikova et al.(2014)]{2014ApJ...786..125S} Starikova, S., Jones, 
C., Forman, W.~R., et al.\ 2014, \apj, 786, 125 


\bibitem[Tonry 
\& Davis(1979)]{1979AJ.....84.1511T} Tonry, J., \& Davis, M.\ 1979, \aj, 84, 1511 

\bibitem[Utsumi et al.(2014)]{2014ApJ...786...93U} Utsumi, Y., Miyazaki, 
S., Geller, M.~J., et al.\ 2014, \apj, 786, 93 

\bibitem[Van Waerbeke et al.(2013)]{2013MNRAS.433.3373V} Van Waerbeke, L., 
Benjamin, J., Erben, T., et al.\ 2013, \mnras, 433, 3373 


\bibitem[Vergani et 
al.(2008)]{2008A&A...487...89V} Vergani, D., Scodeggio, M., Pozzetti, L., et al.\ 2008, \aap, 487, 89 


\bibitem[Viola et al.(2015)]{2015MNRAS.452.3529V} Viola, M., Cacciato, M., 
Brouwer, M., et al.\ 2015, \mnras, 452, 3529 



\bibitem[Westra et al.(2010)]{2010ApJ...708..534W} Westra, E., Geller, 
M.~J., Kurtz, M.~J., Fabricant, D.~G., 
\& Dell'Antonio, I.\ 2010, \apj, 708, 534 





\bibitem[Wittman et al.(2006)]{2006ApJ...643..128W} Wittman, D., 
Dell'Antonio, I.~P., Hughes, J.~P., et al.\ 2006, \apj, 643, 128 







\bibitem[Zahid et al.(2013)]{2013ApJ...771L..19Z} Zahid, H.~J., Geller, 
M.~J., Kewley, L.~J., et al.\ 2013, \apjl, 771, L19 


\bibitem[Zahid et al.(2014)]{2014arXiv1404.7526Z} Zahid,H.~J., Dima, G., 
Kudritzki, R., et al.\ 2014, arXiv:1404.7526 


\end{thebibliography}
\end{document}